\documentclass[amsmath,amssymb,aps,eqsecnum]{revtex4}
\usepackage{amsmath}
\usepackage[dvips]{graphicx}
 \usepackage{bm,bbm}
\usepackage{epsfig}
\usepackage{color}
\usepackage{amssymb,mathtools,amsthm}

 \usepackage{bm}

\usepackage[dvipsnames]{xcolor}

\newcommand{\C}{\mathbb{C}}

\begin{document}
\title{Multipartite entanglement}
\author{Pawe{\l} Horodecki$^{1,2}$, {\L}ukasz Rudnicki$^{1}$  and Karol {\.Z}yczkowski$^{3,4}$}

\affiliation{$^1$International Centre for Theory of Quantum Technologies, University of Gdansk, 80-308 Gda{\'n}sk, Poland}

\affiliation{$^2$Faculty of Applied Physics and Mathematics, Technical University of Gdansk, 80-952 Gda{\'n}sk, Poland}

\affiliation {$^3$Institute of Theoretical Physics, Jagiellonian University,
30-348 Krak{\'o}w, Poland}
\affiliation{$^4$Center for Theoretical Physics, 
     Polish Academy of Sciences, 02-668 Warsaw, Poland}

 \date{July 31, 2024} 

\begin{abstract}
In this contribution
we present a concise introduction to quantum entanglement in multipartite systems. After a brief comparison between bipartite systems and the simplest non-trivial multipartite scenario involving three parties, we review mathematically rigorous definitions of separability and entanglement between several subsystems, as well as their transformations and measures. 
\end{abstract}

\maketitle


\section{Introduction}

A seminal paper from 1935 by  Einstein, Podolsky and Rosen
demonstrated that quantum theory admits 
particular states describing a bipartite physical system, 
which display non-classical correlations between the outcomes 
of measurements performed on two subsystems separated in space.
Their thought experiment was aimed to show that 
`the quantum-mechanical description of physical reality given by wave functions is not complete' \cite{EPR35}.
The very same year Erwin Schr{\"o}dinger analyzed
`probability relations between separated systems',
introducing the term {\sl quantum entanglement} 
and calling it {\sl the characteristic trait of quantum mechanics}
\cite{Sc35}.  His famous  thought experiment, called {\sl Schr{\"o}dinger's cat}, 
illustrates a paradox related to quantum superposition and quantum entanglement.

This topic was revived in the 1960s, as John Bell
introduced a test, nowadays known as the {\sl Bell inequality} \cite{Be64},
which was shown to be satisfied by any theory obeying  {\sl local realism}. 
Results of experiments by Freedman and Clauser \cite{FC72},
and Aspect et al. \cite{AGR82}, published in 1972 and 1982 respectively,
proved that Bell inequalities can be violated, 
in agreement with predictions of quantum theory. 
Further progress in theory of quantum entanglement (comprehensively reviewed in \cite{HHHH09})
was marked by the works of 
Primas \cite{Pri84}
and Werner \cite{We89}, who generalized entanglement for the space of density matrices,
and by the work of Ekert, who showed direct applications of quantum entanglement
for quantum cryptographic schemes \cite{Ek91}.

The field of multipartite entanglement was initiated in the seminal paper of Svetlichny \cite{Sv87} and popularised by
Greenberger, Horne and Zeilinger \cite{GHZ89}. A concise mathematical formulation of the last phenomenon and its major features is the subject of our contribution. To this end, we begin in Sec. \ref{sec:scene} with a quantifiable intuition concerning differences between systems involving two and three parties, the latter being the first non-trivial multipartite scenario. Then, in Sec. \ref{sec:multi_mixed} we continue with a theoretic description of the main subject.

\section{Setting the scene} 
\label{sec:scene}

To present the topic of quantum entanglement we will use the standard toolbox
of quantum mechanics \cite{BZ06}. The term {\sl state} represents a
mathematical object used to calculate probabilities of measurement outcomes.
In the classical theory one works with probability vectors, $p=\{p_1, \dots, p_d\}$,
such that $p_i\ge 0$ and $\sum_{i=1}^d p_i=1$. The fixed natural number $d$,
assumed to be finite, encodes the number of distinguishable events.
The degree of mixedness of the vector $p$ is described by its {\sl purity}, 
$R=\sum_{i=1}^d  p_i^2 \in [1/d,1]$, or its {\sl Shannon entropy}, $S(p)=-\sum_{i=1}^d  p_i \log p_i \in [0, \ln d]$ (the base of the logarithm is a matter of convention in general, however in classical and quantum information theory it is usually chosen to be binary which harmonises with the choice of the bit as a classical information unit).
 
In quantum theory 
one introduces the notion of a {\sl pure state}:
 a vector $|\psi\rangle$ from a $d$-dimensional complex Hilbert space ${\mathcal H}$.
It is convenient to assume the normalization,
 $||\psi||^2=\langle \psi|\psi\rangle=1$,
  and to identify all vectors differing by a complex  phase,
   $|\psi\rangle \sim e^{i \alpha}|\psi\rangle$,  with $\alpha\in [0, 2 \pi]$. 
In the case of a two-level system,  $d=2$, also called a {\sl qubit},
the space of all pure states forms the {\sl Bloch sphere},  $S^2=\C P^{1}$.
For any higher dimension $d$ it forms a complex projective space, $\C P^{d-1}$,
of $2(d-1)$ real dimensions \cite{BBZ02}. The set of pure quantum states is continuous,
in contrast to the discrete set of classical pure states --- the corners of the
probability simplex $\Delta_d$.

\smallskip
A Hermitian operator $P_{\psi}= |\psi \rangle \langle \psi|=P_{\psi}^2$ 
is a projection operator onto a pure state  $|\psi \rangle$. 
Any convex combination of such projectors,
$\rho =\sum_{j=1}^k  q_j  |\psi_j \rangle \langle \psi_j|$,
forms a mixture of pure states, 
where $q$ represents a probability vector of an arbitrary length $k$.
Such a mixture, called a {\sl density matrix} or a {\sl mixed state}, 
can also be defined as a complex Hermitian matrix of order $d$,
which is positive semi-definite, $\rho=\rho^{\dagger}\ge 0$,
and normalized, 
${\rm Tr}\rho=1$.
 
We will 
use an example of pure states to highlight major features of multipartite entanglement. While extending the above mathematical toolbox to cover quantum entanglement in both fundamental and operational aspects, we start with "most basic" bipartite systems and then compare them with "less trivial" tri-partite scenario.

\medskip

\subsection{Bipartite entanglement of pure states }
\label{sec:bipart}

Consider a physical system with an internal structure,
so that one can identify its two subsystems $A$ and $B$, for simplicity both assumed to be  of dimension $d$,
described in Hilbert spaces  ${\mathcal H}_A$ and ${\mathcal H}_B$ respectively.
Then, such a {\sl bipartite} system $AB$ of size $d^2$ can be represented by
a quantum state from the composite Hilbert space
with a tensor product structure,
 ${\mathcal H}_{AB}= {\mathcal H}_A \otimes {\mathcal H}_B$.
 Assuming that both $A$ and $B$ are well defined and the 
 above splitting  of ${\mathcal H}_{AB}$ is fixed,
one can introduce the following key notions \cite{Be64,HHHH09}:
\smallskip

\begin{enumerate}
  \item[a)]  A bipartite pure quantum  state
$|\psi_{AB}\rangle \in   {\mathcal H}_{AB}$
 is called {\sl separable}, if it has the product form,
$|\psi_{AB}\rangle =|\phi_A\rangle \otimes |\phi_B\rangle$,
where  $|\phi_A\rangle  \in {\mathcal H}_A$ and  $|\phi_B\rangle  \in {\mathcal H}_B$.
\item[b)] A bipartite pure quantum  state $|\psi_{AB}\rangle$ 
 is called {\sl entangled}, if it is not separable.  
\end{enumerate}
Note that the above definitions  do not depend on the choice of the
local bases in both subspaces,
however they fundamentally depend on the splitting  of ${\mathcal H}_{AB}$
into ${\mathcal H}_A $ and ${\mathcal H}_B$. 
Thus, the notion of {\sl entanglement} is invariant with respect to
{\sl local unitary} (LU) transformations, $U_{\rm loc} \in \mathcal{U}(d) \otimes \mathcal{U}(d)$.
It is convenient to introduce a broader class of transformations,
called {\sl local operations and classical communication} (LOCC),
in which all local  quantum operations, including measurements
performed on any subsystem, are taken into account.
With the help of classical communication both parties can exchange 
classical information and introduce classical
correlations between subsystems \cite{CLMOW14}.
Therefore, LOCC transformations cannot increase entanglement between subsystems, but they can preserve it or decrease it.

\bigskip

Let $\{ |i\rangle\}_{i=1}^d$ denote an orthonormal basis, 
so that $\langle i|j\rangle =  \delta_{ij}$. Any pure state can be decomposed in this basis,
$|\psi\rangle = \sum_{i=1}^d c_i |i\rangle$, and the normalization condition
reads $\sum_{i=1}^d |c_i|^2=1$.
 In a bipartite, $d \times d$ system
${\cal H}_A \otimes {\cal H}_B$, one introduces a product basis,
$|i,j\rangle\ = |i\rangle_A \otimes |j\rangle_B$
with  $i,j=1,\dots, d$ such that $\langle i,j| i'j'\rangle =  \delta_{ii'}\delta_{jj'}$.
 Then, any pure state of two subsystems with $d$ levels each can be written as
\begin{equation}
\label{pure_bi1}
|\psi_{AB} \rangle =\sum_{i=1}^d \sum_{j=1}^d  G_{ij} \;
  |i, j\rangle . 
\end{equation}
Using the singular value decomposition of the matrix of coefficients,
 $G=UDV^{\dagger}$, where $U$ and $V$ are unitary 
and $D={\rm diag} (\sqrt{\lambda_1}, \dots, \sqrt{\lambda_d})$ is diagonal with real non-negative entries,
the same state can be written in its {\sl Schmidt form},
\begin{equation}
\label{pure_bi2}
|\psi_{AB}\rangle =  ( U \otimes V^{\dagger}) \sum_{i=1}^d  \sqrt{\lambda_i} \;
   |i\rangle_A \otimes |i\rangle_B .
\end{equation}
The normalization condition, $\langle \psi_{AB} |\psi_{AB} \rangle=1$,
implies ${\rm Tr}GG^{\dagger}=1$ and in turn $\sum_{i=1}^d \lambda_i=1$,
so the Schmidt vector $\lambda$ forms a $d$-point probability distribution,
which  characterizes entanglement properties of the state.
The state is separable, if matrix $G$ is of rank one, so 
there exists a single, non-zero Schmidt coefficient, $\lambda_{\rm max}=1$,
and the Shannon entropy vanishes, $S(\lambda)=0$. The other extreme case of a flat vector, 
$\lambda_i=1/d$, leads to the maximal entropy, $S(\lambda)=\ln d$,
which distinguishes the class of maximally entangled Bell states,
$|\psi^{+,d} \rangle= \frac{1}{\sqrt{d}} \sum_{i=1}^d |i\,i\rangle$,
where $|i\, i\rangle= |i\rangle_A \otimes |i\rangle_B.$

Any physical system, described by a (mixed) state $\rho$ can be always {\sl extended} 
by adding an auxiliary subsystem $\omega$ to produce a tensor product, 
$\rho \to \sigma=\rho \otimes \omega$.
The inverse process of generating a reduced state is obtained by a {\sl partial trace}.
Consider a bipartite state $\sigma$ of size $d^2$ and expand it in a product basis,
$ \sigma_{\stackrel{\scriptstyle m \mu}{n \nu}}
= \langle m, \mu |\sigma| n, \nu \rangle$,
where Latin and Greek indices represent subsystems $A$ and $B$, respectively.
The trace of this bipartite state is by definition given by a double sum, 
${\rm Tr}\sigma=\sum_{m,\mu=1}^d  \sigma_{\stackrel{\scriptstyle m \mu}{m\mu}}$.
The partial trace, given by a sum over a single index,
produces  both {\sl reduced} states,
$\sigma^A_{mn}=\sum_{\mu=1}^d  \sigma_{\stackrel{\scriptstyle m \mu}{n \mu}}$
and
$\sigma^B_{\mu,\nu}=\sum_{m=1}^d  \sigma_{\stackrel{\scriptstyle m \mu}{m \nu}}$,
also written
$\sigma_A={\rm Tr}_B\sigma$ and $\sigma_B={\rm Tr}_A\sigma$.

Taking partial traces of the projector on a bipartite pure state (\ref{pure_bi2})
with respect to a subsystem $A$ or $B$
one obtains mixed states $\rho_B=G^{\dagger}G$ and $\rho_A=GG^{\dagger}$,
both having the same spectrum $\lambda=\{\lambda_1, \dots, \lambda_d\}$.
The degree of entanglement of any bipartite  pure state  $|\psi_{AB}\rangle$
can be thus characterized by the degree of mixedness of both reduced states!
The latter can be measured by the {\sl entanglement entropy},
given by the von Neumann entropy of the partial trace, 
and equal to  the Shannon entropy of the Schmidt vector, 
$E(|\psi_{AB}\rangle)=S(\rho_A)=S(\rho_B)=S(\lambda)$.
The state  $|\psi_{AB}\rangle$ is separable iff its entanglement entropy vanishes, 
$E(|\psi_{AB}\rangle)=0$, and it is  called {\sl maximally entangled},
if the Schmidt vector is flat, $\lambda_i=1/d$, so the entropy is maximal, $E=\ln d$, 
and the matrix $G$ is unitary up to rescaling, $GG^{\dagger}={\mathbbm 1}/d$.
A simple function of the purity of the reduced matrix,
$\tau=2(1-\sum_{i=1}^d \lambda_i^2)$, normalized to take values in $[0,1]$,
 is called {\sl tangle}, and its square root, $C=\sqrt{\tau}$, is called
 {\sl concurrence}. In the case of a two-qubit system one has $1=\lambda_1+\lambda_2$,
 so that the tangle is related to the determinant, 
 \begin{equation}
 \label{tau}
 \tau=4\lambda_1\lambda_2=4|{\rm det} G|^2.
 \end{equation}
It is convenient to organize the Schmidt vector in a non-increasing order, 
$\lambda_1\ge \lambda_2 \ge\dots \ge \lambda_d\ge 0$. It is then possible to compare any two bipartite pure states 
$|\psi\rangle$ and $|\phi\rangle$ with Schmidt vectors $\lambda$ and $\mu$, respectively,
by using the {\sl majorization} relation. 
If $\sum_{i=1}^k\lambda_i \ge \sum_{i=1}^k \mu_i$ for any $k=1,\dots, d-1$,
we say that the vector $\lambda$ majorizes $\mu$, written $\lambda \prec \mu$,
and consequently  $|\psi\rangle  \prec |\phi\rangle$.
This majorization relation introduces a partial order into the set of  pure states:
due to the theorem of Nielsen \cite{Ni99} 
 a state $|\psi\rangle$ can be transformed by LOCC into 
$ |\phi\rangle$ iff  the relation $|\psi\rangle  \prec |\phi\rangle$ holds.

We will also use a broader class of operations, in which 
a bipartite pure state $|\psi\rangle$ is transformed with some {\sl probability}
into a non-normalized state $|\phi\rangle$. Such operations, called {\sl stochastic LOCC} or SLOCC,
are possible if there exist two invertible matrices $L_1$ and $L_2$
such that $|\phi\rangle = L_1 \otimes L_2 |\psi\rangle$.
Since in this relation the normalization does not play any role, we may assume that matrices $L_j$
belong to the special linear group $SL(d)$ of matrices with determinant equal to unity.
The SLOCC transformations can decrease the entanglement entropy of
 a bipartite state $|\psi_{AB}\rangle$, but will preserve its {\sl Schmidt rank},
 defined by the number of positive components of the Schmidt vector $\lambda$.
Such quantities which do not  decrease under any local manipulation of the system
are called {\sl entanglement monotones} \cite{Vi99,Vi00}.

\subsection{Tripartite entanglement of pure states}
\label{sec:multi_pure}
In analogy with the bipartite case (\ref{pure_bi1}),
any pure state of a tripartite state 
can be represented in a product basis,
\begin{equation}
|\psi_{ABC} \rangle =\sum_{i=1}^d \sum_{j=1}^d \sum_{k=1}^d  T_{ijk} 
  |i\rangle_A \otimes |j\rangle_B \otimes  |k\rangle_C . 
\label{three-tensor} 
\end{equation}
However,  a dimension-counting argument implies that 
a general tripartite state 
$|\psi_{ABC} \rangle$ 
{\sl cannot} be written in the form \cite{CHS00},
\begin{equation}
|\psi_{ABC}\rangle =  ( U_A \otimes U_B \otimes U_C ) \sum_{i=1}^d  \sqrt{\lambda_i} \;
   |i\rangle_A \otimes |i\rangle_B \otimes |i\rangle_C \ ,
\label{threeB} 
\end{equation}
analogous to the Schmidt decomposition (\ref{pure_bi2}). 
In short,  algebraic operations on matrices are
much simpler than their analogues applied to tensors.
Singular value decomposition allows us to transform
any matrix $G_{ij}$ to the diagonal form
by two unitary rotations, but even three unitaries are not sufficient 
\cite{SA14}  to bring an arbitrary
tensor $T_{ijk}$ to the form (\ref{threeB}).
Multipartite entanglement is, therefore,
 more sophisticated than the bipartite case 
and it has a rich phenomenology already for pure states.
If one considers the number $N$  of parties in a quantum composite system,
then three is much more than two, and also four is more than three...

\medskip

\begin{figure} [htbp]
   \begin{center} 
 \includegraphics[width=7.1cm,angle=0]{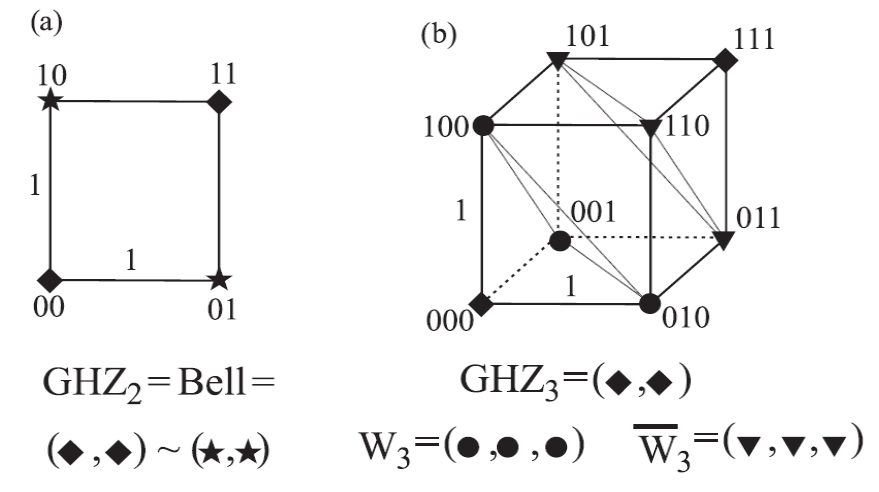}
\vskip -0.5cm
\caption{Distinguished states for a) two qubits:
superposition of states corresponding to
two points at an edge forms a separable state,
two points at a diagonal of the square correspond to a Bell state;
b) three-qubit systems: 
two points at a diagonal    of a face of the cube denote biseparable states, 
two points at the diagonal of the cube represent GHZ states,
while three points forming an equilateral triangle
correspond to a W state. To analyze a four-qubit system one needs a hypercube.}
 \label{fig_cube}
   \end{center} 
 \end{figure}

\begin{figure} [htbp]
   \begin{center} 
a) 
 \includegraphics[width=4.5cm,angle=0]{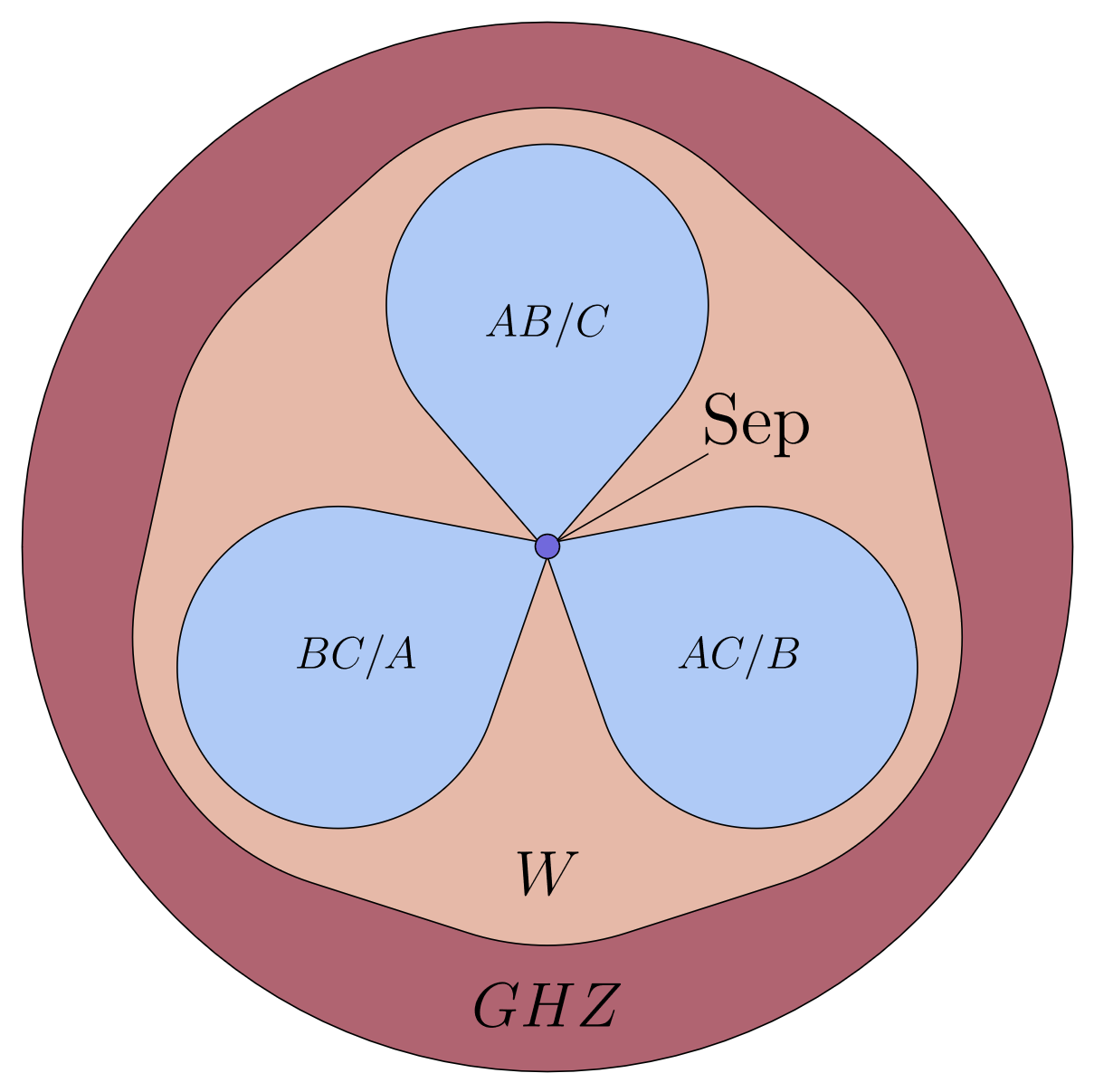} \ \  \ \ \ \  b)
 \includegraphics[width=5.1cm,angle=0]{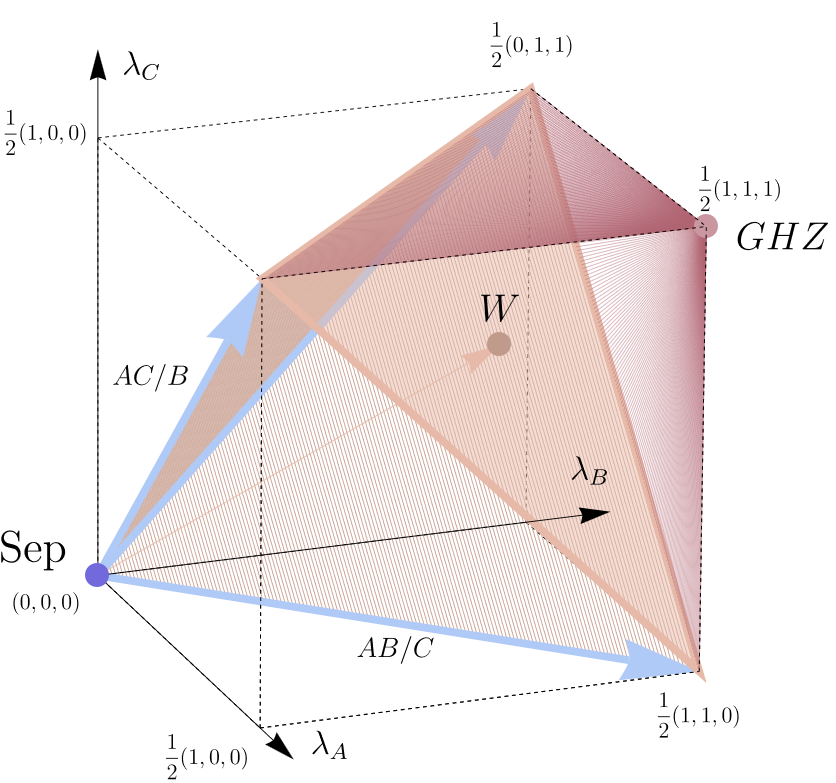}
\vskip -0.3cm
\caption{Set of three-qubit pure states including
separable states, biseparable states,  states equivalent with respect to 
SLOCC to $|W\rangle$ and $|GHZ\rangle$:
 a) an 'artist's sketch' (based on \cite{ABLS01}),
  b) actual positioning in the 
  {\sl entanglement polytope} \cite{HSS03,SWK13,WDGC13}
  spanned by the smaller eigenvalues
     $\lambda_A, \lambda_B$ and $\lambda_C$ of three single-partite reduced states.}
      \label{fig_3ab}
     \end{center} 
 \end{figure}

In the case of two qubits we have identified
 the maximally entangled {\sl Bell state},
$|\psi^{+,2}\rangle= \frac{1}{\sqrt{2}} ( |00\rangle +  |11\rangle)$ 
(where the superscript $2$ refers to the dimension of local Hilbert space).
In the three-qubit system it will be convenient to distinguish
 two particular states -- 
 see Figs. \ref{fig_cube}b
 and \ref{fig_3ab}. 
\begin{equation}
|GHZ\rangle =   \frac{1}{\sqrt{2}} ( |000\rangle +  |111\rangle) 
\quad  \quad {\rm and} \quad \quad
|W\rangle =   \frac{1}{\sqrt{3}}( |001\rangle +  |010\rangle  + |100\rangle).
\label{GHZ_W} 
\end{equation}
The first state, $|GHZ\rangle$, 
became popular due to the paper
of Greenberger, Horne, and Zeilinger \cite{GHZ89},
in which tripartite quantum correlations were analyzed, 
but it appeared earlier in the work of Svetlichny  \cite{Sv87}.
Such a three--qubit GHZ state was realized in an experiment in 1999 \cite{BPDWZ99}.  
The second state, $|W\rangle$,  was used in \cite{DVC00} to show
that  `three qubits can be entangled in two inequivalent ways'.

Both states specified in Eq. (\ref{GHZ_W}) 
are not locally equivalent.
To verify LU equivalence for any two-qubit pure states it is sufficient
to check, if the purity of the partial traces of a first state 
matches analogous quantity evaluated for the second one. 
However, in the case of three qubits one needs to compare several
 invariants  \cite{LPS99,Su01,BL01} 
of the composed group $\mathcal{U}(2)^{\otimes 3}$.

The first LU invariant of order two is just the norm of the state  
$I_1=\langle \psi | \psi \rangle$, which is fixed by the normalization.
There are three invariants of order four given by  purities of the three single--party reductions,
\begin{equation}
\label{I2_I4} 
I_2= {\rm Tr} \rho_A^2, \quad \quad
I_3= {\rm Tr} \rho_B^2, \quad \quad
I_4= {\rm Tr} \rho_C^2, 
\end{equation}
where $\rho_A={\rm Tr}_{BC} |\psi_{ABC}\rangle  \langle \psi_{ABC}|$, etc.
By construction, one has $1/2 \le I_i \le 1$  for $i=2,3,4$.
The fifth invariant  $I_5$, called the   {\it Kempe invariant} \cite{Ke99},
is of order six 
and relates the third moment of partial traces
and two-party reduced states,
\begin{equation} 
\label{I5}
I_5 = 3 {\rm Tr}[(\rho_A \otimes \rho_B)\; \rho_{AB}] 
- {\rm Tr}\rho_A^3  -{\rm Tr}\rho_B^3.
\end{equation}
It can be expressed in two equivalent ways 
by exchanging subsystems $B \leftrightarrow C$
and later $A \leftrightarrow B$.
Thus, this invariant is symmetric with respect to exchange of the subsystems, and 
it satisfies $2/9 \le I_5 \le 1$, with the minimum attained by the $W$ state.

A sixth invariant $I_6$  is of order eight.
It was identified by Coffman, Kundu and Wootters \cite{CKW00}
and can be 
expressed by the 
 {\it hyperdeterminant} $\rm {Det}_3$  of the $3$-tensor $T_{ijk}$,
 introduced by Cayley  \cite{Ca45} already in 1845,
\begin{equation}
\label{I6}
 I_6 =    4 |{\rm Det}_3(T)|^2 = \frac{1}{4} \tau_3^2\ ,   
 \end{equation}
which reads \cite{GKZ94}, 
\begin{eqnarray} 
\label{det33}
{\rm Det}_3(T) = 
  [T_{000}^2 T_{111}^2 + T_{001}^2 T_{110}^2 
 + T_{010}^2 T_{101}^2  + T_{100}^2 T_{011}^2] 
   - 2 \bigl[ T_{000}  T_{111} ( T_{011} T_{100}+ T_{101} 
	T_{010} +T_{110} T_{001} )  \nonumber \\ 
  + T_{011} T_{100} ( T_{101} T_{010}+ T_{110} T_{001} ) +                
  T_{101} T_{010} T_{110} T_{001} \bigr]  
       +  4 \bigl[  T_{000} T_{110} T_{101} T_{011} + 
	T_{111}  T_{001} T_{010} T_{100} \bigr] .
\end{eqnarray}
Note that this expression  consists of three terms
  with the property, that
their “center of mass” coincides with that of the 
 underlying cube \cite{CKW00}, in analogy to the 
 determinant of a matrix of order two.
To demonstrate that all the invariants $I_1, \dots, I_6$
 are independent one can show \cite{Su01}
that their gradients are linearly independent at some point.
However, the set of six invariants $I_i$ is not sufficient to single out a local orbit,
as $I_i(|\psi\rangle ) = I_i(|\psi\rangle^*)$ for all of them. 
The problem of deciding when two given multiqubit states can 
be connected by local unitaries was analyzed in \cite{Kra10}. 

Note that in the case of a two-qubit pure state, 
the determinant of the matrix  $G$ defining the state  (\ref{pure_bi1})
determines a measure of entanglement called tangle, $\tau=4|{\rm det} G|^2$.
In a similar way, for a three-qubit state,  the  hyperdeterminant
(\ref{det33}) is related to the entanglement measure $\tau_3$
called {\sl three--tangle} \cite{CKW00,DVC00}. To introduce this notion one  analyzes first the entanglement
between a single party $A$ and the composite system $BC$, 
written $A|BC$ and compares it with 
the pairwise entanglement $A|B$ and $A|C$.
The following  {\it monogamy relation}
concerning tangle, equal to squared concurrence, $\tau=C^2$
 was established in \cite{CKW00},
\begin{equation}
\tau_{A|BC} \ge \tau_{A|B}  + \tau_{A|C} \ge 0 .
\label{monogamy}
\end{equation}
Here $\tau_{A|B}$ denotes the tangle of the two--qubit reduced state, 
$\rho_{AB}={\rm Tr}_C \rho_{ABC}$,
while $\tau_{A|BC}$ represents the tangle 
between part $A$ and the composite system $BC$.
Although the subsystem $A$ can be simultaneously 
entangled with the remaining subsystems $B$ and $C$,  
the "sum of these two types of entanglement"
cannot exceed the entanglement between $A$ and $BC$.
Entanglement in a three--qubit pure state,
$\rho_{ABC}=|\psi_{ABC}\rangle\langle \psi_{ABC}|$,
can be thus described by the  quantity $\tau_1$ called {\sl one-tangle},
obtained by  averaging over three possible splittings,
\begin{equation}
\tau_1\bigl( |\psi_{ABC}\rangle \bigr) 
\equiv \frac{1}{3} \bigl (\tau_{A|BC} + \tau_{B|AC} + \tau_{C|AB}\bigr) \ .
\label{tau1}
\end{equation}
Related quantity, two-tangle,  characterizes
average entanglement contained in two-partite reductions, 
\begin{equation}
\tau_2\bigl( |\psi_{ABC}\rangle \bigr) \equiv \frac{1}{3} \bigl (\tau_{A|B} + \tau_{B|C} + \tau_{C|A}\bigr) 
\ .
\label{tau2}
\end{equation}

\noindent Both quantities $\tau_1, \tau_2$ are non-negative
by construction and can be evaluated analytically by the formula of
 Wootters \cite{Wo98}.
It is simple to check that $\tau_1$ achieves its maximum for the GHZ state,
$0 \le \tau_1\bigl( |\psi_{ABC}\rangle \bigr)  \le 
\tau_1\bigl( |GHZ \rangle \bigr) =1$,
while two-tangle  is maximized by the $W$ state,
$ 0 \le \tau_2\bigl( |\psi_{ABC}\rangle \bigr)  \le 
\tau_2\bigl( |W \rangle \bigr) =4/9.$

\begin{table}[h]
\caption{Exemplary three-qubit states $|\psi_{ABC}\rangle$,
 their LU invariants $I_1$--$I_6$, 
tangles $\tau_1$--$\tau_3$ and ranks $r_X$ of single--partite reductions,
which are invariant with respect to SLOCC transformations;
 $|\psi^+_{BC}\rangle$ denotes the Bell state in the $BC$ subspace.
}
 \smallskip
\hskip -0.5cm
{\renewcommand{\arraystretch}{1.21}
\begin{tabular}
[c]{c|cccccc|ccc|ccc|c} 
\hline \hline
State & $I_1$ &  $I_2$  & $I_3$ & $I_4$  & $I_5$ & $I_6$ & $\tau_1$ &  $\tau_2$ & $\tau_3$ 
& $r_A$ & $r_B$ & $r_C$ & entanglement \\
  \hline \hline
  $  |\phi_A\rangle \otimes |\phi_{B}\rangle \otimes |\phi_C\rangle $                          
                                 & $1$ & $1$ & $1$ & $1$ &  $1$ & $0$ & $0$ & $0$ & $0$ & $1$ & $1$ & $1$ & none \\
$|\phi_A\rangle \otimes |\psi^+_{BC}\rangle $ & $1$ & $1$ & $1/2$ & $1/2$ &  $1/4$ & $0$ & $2/3$ & $1/3$ & $0$ & $1$ & $2$ & $2$ & bipartite \\
$|\phi_B\rangle \otimes |\psi^+_{AC}\rangle $ & $1$ & $1/2$ & $1$ & $1/2$ &  $1/4$ & $0$ & $2/3$ & $1/3$ & $0$ 
& $2$ & $1$ & $2$ & bipartite\\
$|\phi_C\rangle \otimes |\psi^+_{AB}\rangle $ & $1$ & $1/2$ & $1/2$ & $1$ &  $1/4$ & $0$ & $2/3$ & $1/3$ & $0$
& $2$ & $2$ & $1$ & bipartite \\
                               $|W\rangle $ & $1$ & $5/9$ & $5/9$ & $5/9$ &  $2/9$ & $0$ & $8/9$ & $4/9$ & $0$ 
                               & $2$ & $2$ & $2$ & triple bipartite
                               \\
                             $|GHZ\rangle $ & $1$ & $1/2$ & $1/2$ & $1/2$ &  $1/4$ & $1/4$ & $1$ & $0$ & $1$
                             & $2$ & $2$ & $2$ & global tripartite \\
\hline \hline
\end{tabular}
}
\label{tab:3qub3}
\end{table}
\smallskip

The most important measure characterizing the {\it global entanglement},  
based on the monogamy relation (\ref{monogamy}),
is  therefore the {\it $3$--tangle} \cite{CKW00},
\begin{equation}
\tau_3\bigl( |\psi_{ABC}\rangle \bigr) \equiv  
\tau_{A|BC} - \tau_{A|B} - \tau_{A|C}. 
\label{tau3}
\end{equation}
It is invariant with respect to a permutation of the subsystems.
One uses in parallel the name {\it residual entanglement},
as $\tau_3$ characterizes the fraction of entanglement which cannot be 
described by any two-body measures. 
Furthermore, 3-tangle is invariant under the action of 
the group $G_L=GL(2,{\mathbb C})^{\otimes 3}$.
As $\tau_3$ vanishes for the $W$ state 
and is equal to  unity for the GHZ state  -- see Table \ref{tab:3qub3} -- 
it can be used to distinguish the class of states
accessible by SLOCC operations from $|W\rangle$
and the states accessible from the
$|GHZ\rangle$ state.
All entanglement measures are invariant under local unitaries,
so the  tangles $\tau_i$  are functions of the unitary invariants $I_j$. Introducing 
the average invariant of order four, $I_{\rm av}=(I_2+I_3+I_4)/3$,
it is easy to check that  $\tau_1=2(1-I_{\rm av})$,
and  $\tau_2=1-I_{\rm av} - 2I_6$, while $\tau_3=2 \sqrt{I_6}$.

The Schmidt decomposition implies that any two-qubit state
can be brought by a local unitary to  the form,
 $|\psi_{\theta}\rangle = \cos\theta |00\rangle + \sin\theta |11\rangle$,
as both unitary rotations in Eq.~(\ref{pure_bi2}) allow us to  set a single component to zero.
In a similar way, applying a local operation $U\in \mathcal{U}(2)^{\otimes 3}$
to any three-qubit pure state
one can set three components out of eight to zero \cite{CHS00},
and bring the state  to its canonical form consisting of $8-3=5$ components \cite{AACJLT00,AAJ01},
\begin{equation}
|\psi\rangle = r_0e^{i\theta}|000\rangle +r_1|100\rangle +
                 r_2|010\rangle +r_3|001\rangle +r_4|111\rangle \ ,
\label{acinB}
\end{equation}
and determined by five parameters:
four real amplitudes $r_j$ and an arbitrary phase $\theta$.
It is known that a generic pure state of three qubits is completely determined by
its two-particle reduced density matrices \cite{LPW02,SWK13}.
Updated information on pure states
entanglement in 
many body systems can be found in a
readable review \cite{SMLR24}.

\section{Multipartite entanglement of mixed states} 
\label{sec:multi_mixed}
So far 
we were mostly concerned with pure states. However, in the case of density matrices, useful for systems
interacting with an environment or being subject to a measurement,
already for the bipartite scenario one needs a more sophisticated definition of separability.
The one proposed by Werner \cite{We89} requires that 
a separable state has a product structure, corresponding to independent probabilistic variables,
 or can be obtained by a mixture of such states:
 
\smallskip

\begin{enumerate}
  \item[a')]  A bipartite mixed state
$\rho^{AB}$ acting on ${\mathcal H}_A \otimes {\mathcal H}_B$
 is called {\sl separable},  if it can be represented as a convex combination of
 product states,  $\rho_j^A \otimes \rho_j^B$,
 \begin{equation}
 \label{sep_mixed0}
 \rho^{AB}_{\rm sep} \ = \ \sum_{j=1}^k   q_j  \;  \rho_j^A \otimes \rho_j^B,
 \end{equation} 
 \noindent
 where $q$ is a probability vector of a finite length $k$.
 
\item[b')]  A bipartite mixed state
$\rho^{AB}$  
 is called {\sl entangled}, if it is not separable.  
\end{enumerate}
 In the case of pure states, which cannot be represented by a mixture of other pure states,
 both definitions of entanglement 
 are  consistent.

Within the bipartite scenario, Eq. (\ref{sep_mixed0}) contains an ambiguity related with the convex decomposition --- given a separable state one does not know the form of a suitable ensemble of the product states. However, since one deals with two subsystems only, it is clear that the (mixed) product states are of the form $\rho^A\otimes \rho^B$. In the multipartite setting, also the latter becomes ambiguous (cf. the previous section and Table \ref{tab:3qub3} therein). Therefore, 
we begin discussing the "landscape" of multipartite separability 
and quantum entanglement. In other words, considering $N$ subsystems we establish counterparts of definitions a), b) and a'), b'), 
such that
for $N=2$ parties
their union boils down to the former definitions. 

We start from a notion of an $N$-partite system which is described in the Hilbert space 
\begin{equation}\label{spl1}
  \mathcal{H}_{A_1 \cdots A_N}=\bigotimes_{i=1}^N \mathcal{H}_{A_i}.
\end{equation}
As before, we assume that the above splitting is fixed, i.e., Hilbert spaces $\mathcal{H}_{A_i}$ of $N$ "primitive" subsystems have a distinguished (e.g. physical) meaning. We use the word "primitive", because even if a subspace $\mathcal{H}_{A_i}$ can further be split into fine-grained subspaces, we are not going to use this piece information. Otherwise we would rather need to
consider an $M$-partite scenario with $M > N$ at the outset.

We recall that for a bipartite case, after identifying the subspaces $\mathcal{H}_{A}$ and $\mathcal{H}_{B}$ as primitive, the splitting ${\mathcal H}_{AB}={\mathcal H}_A \otimes {\mathcal H}_B$ is unique. Even though quite trivial, it is an important feature from the perspective of the multipartite scenario.
A different splitting ${\mathcal H}_{AB}={\mathcal H}_{A'} \otimes {\mathcal H}_{B'}$ may exist, 
and such a choice corresponds to selecting a different pair of primitive subsystems.

In the multipartite case, the above property does not hold anymore as we can consider instead of (\ref{spl1}) the structure
\begin{equation}\label{spl2}
    \mathcal{H}_{A_1 \cdots A_N}=\bigotimes_{i=1}^{K} \mathcal{H}_{A'_i}.
\end{equation}
Clearly, to maintain the interpretation of $\mathcal{H}_{A_i}$ as primitive subspaces, we need to have $K<N$ and 
 $   \mathcal{H}_{A'_i}=\bigotimes_{j\in S_i} \mathcal{H}_{A_j}$,
where $S_i$ are disjoint sets of indices such that $\bigcup_{i=1}^K S_i = \{1,\ldots,N\}$.
In other words, the \textit{partition} (\ref{spl2}) is \textit{coarser} than (\ref{spl1}), while (\ref{spl1}) is \textit{finer} than (\ref{spl2}). In this way we face a formal notion of the partition $\alpha_K=\left(S_1,\ldots,S_K\right)$ of the Hilbert space $\mathcal{H}_{A_1 \cdots A_N}$. This is the partition of
the set $\{1,\ldots,N\}$ into $K$ disjoint and non-empty subsets \cite{Kpartite}.
 
 The boundary choices $K=1$ and $K=N$ are defined in a unique way:  $\alpha_1=(\{1,\ldots,N\})$ and $\alpha_N=(\{1\},\ldots,\{N\})$, while for $1<K<N$ there exist several, not necessarily equivalent partitions. 
 The partition $\beta_L=\left(Z_1,\ldots,Z_L\right)$ is {\sl finer} than $\alpha_K$ if every $S_i$ from $\alpha_K$ is a union of some subsets $Z_j$. In this case we say that the partition $\alpha_K$ is {\sl coarser} than $\beta_L$. 
 In general, we do not have such a relation between any two partitions. Still, given a fixed partition $\alpha_K$ we can define a set of partitions, here denoted as $\overline\alpha_K$, containing $\alpha_K$ and all its finer partitions \cite{Szilard}. Note that $\alpha_K$ is the maximal element of $\overline\alpha_K$ with respect to the partial order of fine graining, defined above.

With the appropriate notation and terminology we are ready to collect the list of definitions. For the sake of generality, we establish them already for mixed states. As input we use results developed in \cite{Kpartite, Szilard}, 
but the readers can also consult \cite{sepEXT1,sepEXT2,sepEXT3,sepEXT4,sepEXT5,sepEXT6,sepEXT7,sepEXT8,sepEXT9}
\begin{enumerate}
    \item[$c_1$)] An $N$-partite state acting on $\mathcal{H}_{A_1 \cdots A_N}$ is called {\it fully separable} if it can be represented as a convex combination of
 product states,  $\bigotimes_{i=1}^{N}\rho_j^{A_i}$,
 \begin{equation}
 \label{sep_mixed}
\rho^{A_1 \cdots A_N}_{\rm full-sep} =  \ \sum_{j=1}^k   q_j  \;  \bigotimes_{i=1}^{N}\rho_j^{A_i},
 \end{equation} 
 \noindent
 where every $\rho_j^{A_i}$ acts on $\mathcal{H}_{A_i}$ and $q$ is a probability vector of a finite length $k$.
    
    \item[$c_2$)] An $N$-partite state  acting on $\mathcal{H}_{A_1 \cdots A_N}$ is called { \it $\alpha_K$-separable } for $\alpha_K=\left(S_1,\ldots,S_K\right)$ if it can be represented as a convex combination of
 product states,  $\bigotimes_{i=1}^{K}\rho_j^{A'_i}$, in the form 
 $\rho^{A_1 \cdots A_N}_{\alpha_K \rm -sep} =  \sum_{j=1}^k   q_j   \bigotimes_{i=1}^{K}\rho_j^{A'_i}$,
where every $\rho_j^{A'_i}$ acts on $\mathcal{H}_{A'_i}=\bigotimes_{j\in S_i} \mathcal{H}_{A_j}$ and $q$ is a probability vector of a finite length $k$.

    \item[$c_3$)] An $N$-partite state  acting on $\mathcal{H}_{A_1 \cdots A_N}$ is called $\bar\alpha_K$-separable if for some $K>1$ there exists a partition $\alpha_K$ such that the state can be represented as a convex combination of
 product states,  $\bigotimes_{i=1}^{K_m}\rho_j^{A'_{mi}}$, in the form 
 $\rho^{A_1 \cdots A_N}_{\rm part-sep} =  \ \sum_{j=1}^k   q_j  \;  \bigotimes_{i=1}^{K_m}\rho_j^{A'_{mi}}$,
 where every $\rho_j^{A'_{mi}}$ acts on $\mathcal{H}_{A'_{mi}}=\bigotimes_{j\in S_{mi}} \mathcal{H}_{A_j}$, all partitions $\left(S_{m1},\ldots,S_{mK_m}\right)$ belong to the set $\overline\alpha_K$, and $q$ is a probability vector of a finite length $k$.
\item[$c_4$)] An $N$-partite state  acting on $\mathcal{H}_{A_1 \cdots A_N}$ is called  $K$-separable (for $K>1$) if it can be represented as a convex combination of $\bar\alpha_{K_n}$-separable states, each associated with a partition $\alpha_{K_n}$, such that $\min_n K_n=K$. For $K=2$ the $K$-separable state is called \footnote{This terminology is actually different from the one used in Ref. \cite{Guehne-Toth-2009}, where {\sl biseparable} states denote these called here $\alpha_{K}$-separable for some chosen bipartition $\alpha_{K}=\left(S_1,S_2\right)$.} biseparable \cite{Guehne-Toth-2009}. Equivalently, biseparability means that {\it any} representation of the state as a convex decomposition of pure states contains some pure state that is {\it not product under any bipartition}.

\item[$c_5$)] An $N$-partite state  acting on $\mathcal{H}_{A_1 \cdots A_N}$ is called genuine multipartite entangled if it is 
not biseparable.
\end{enumerate}
We can trivially observe that for bipartite systems we have a unique choice for $\alpha_2$ equal to the primitive partition. Then, the above definitions $c_1$), $c_2$), $c_3$) and $c_4$) coincide, and boil down to a'). In general, however, the definition of $\bar\alpha_K$-separability is finer than that of $K$-separability. There are $K$-separable states which  are not $\bar\alpha_{K'}$-separable for all $\alpha_{K'}$ such that $K'\geq K$. There are also interesting subtleties relevant for sets of states, observed on the level of $\bar\alpha_{K}$-separability. For example, set operations of intersection and union of $\bar\alpha_{K}$ sets imply the same for the associated sets of pure $\bar\alpha_{K}$-separable states, while for mixed states the connection is weaker, as we just get inclusions of the sets of density matrices \cite{Szilard}. 

There is a notion of network entanglement \cite{network1, network2} which is more restrictive than  $c_5$).
\begin{enumerate}
\item[$c_4^{\textrm{net}}$)] An $N$-partite state  acting on $\mathcal{H}_{A_1 \cdots A_N}$ is called a quantum network state if it can be represented as $\rho^{A_1 \cdots A_N}_{\rm net} =  \sum_{j=1}^k   q_j   \bigotimes_{l=1}^{N}\mathcal{E}_{lj}\left(\bigotimes_{i=1}^{K}\rho_i^{A'_i}\right)$, where every $\rho_j^{A'_i}$ and the vector $q$ are defined as in $c_2$). We further assume $K>1$, while $\mathcal{E}_{lj}$ are local quantum operations which do not admit classical communication.
\end{enumerate}
There exist states which are genuine multipartite entangled, still being network states \cite{network1}.

One of the fundamental differences between bipartite and multipartite case is that biseparability is {\it not preserved under tensoring in general} \cite{Yamasaki-Et-Al-22,Palazuelos-de-Vincente-22}.
If only $\rho$ is not $\alpha_{K=2}$-separable under any specific $\alpha_{K=2}$  partition (class $c_2$ above, this separability is also called {\it partial separability}) then there exists a natural number $n$ such that $\rho^{\otimes n}$ is genuinely multipartite entangled \cite{Palazuelos-de-Vincente-22}. In particular, there are such specific constructions for noisy GHZ states \cite{Yamasaki-Et-Al-22}.

To define the notion of \textit{producibility} \cite{producibility}
it is convenient to write down pure states that satisfy the definitions $c_1$) and $c_4$). Note that items $c_2$) and $c_3$) 
differ from $c_4$) for mixed states only. 
Thus
\begin{itemize}
\item A pure state $\bigotimes_{i=1}^{N}\left|\phi_{A_{i}}\right\rangle$ with any $\left|\phi_{A_{i}}\right\rangle \in \mathcal{H}_{A_i}$ meets condition $c_1$).

\item A pure state $\bigotimes_{i=1}^{K}\left|\phi_{A'_{i}}\right\rangle$ with any $\left|\phi_{A'_{i}}\right\rangle\in \mathcal{H}_{A'_i}=\bigotimes_{j\in S_i} \mathcal{H}_{A_j}$ meets condition  $c_4$).
\end{itemize}
Concerning the second bullet point, there is an additional relevant question which has a trivial answer for the first example, and consequently, for all bipartite states. To phrase it, we shall assume that all states  $\left|\phi_{A'_{i}}\right\rangle$ are  genuine multipartite entangled (otherwise, there is a finer partition with respect to which the state assumes a product form). Then, denoting by $\left|S_{i}\right|$ the number of primitive subsystems in each space $\mathcal{H}_{A'_i}$, equal to the cardinality of each subset of indices $S_i$, we ask about $M\coloneqq\max_{i} \left|S_{i}\right|$. 
The number
$M$ quantifies the maximal number of entangled parties and leads
to the following definitions:

\begin{enumerate}
\item[$c_6$)] An $N$-partite pure state  from  $\mathcal{H}_{A_1 \cdots A_N}$ is called $M$-producible if it is $K$-separable, with respect to some partition $\alpha_K=\left(S_1,\ldots,S_K\right)$, and $M=\max_{i} \left|S_{i}\right|$.

\item[$c_7$)] An $N$-partite pure state from  $\mathcal{H}_{A_1 \cdots A_N}$  is called genuine $(M+1)$-partite entangled if it is not $M$-producible.

\item[$c_8$)] As before, convex combinations of $M$-producible pure states give $M$-producible mixed states, while $(M+1)$-partite entangled mixed states are states that cannot be represented as convex combinations of $M$-producible pure states.
\end{enumerate}

 \begin{figure} [htbp]
   \begin{center} 
 \includegraphics[width=13cm,angle=0]{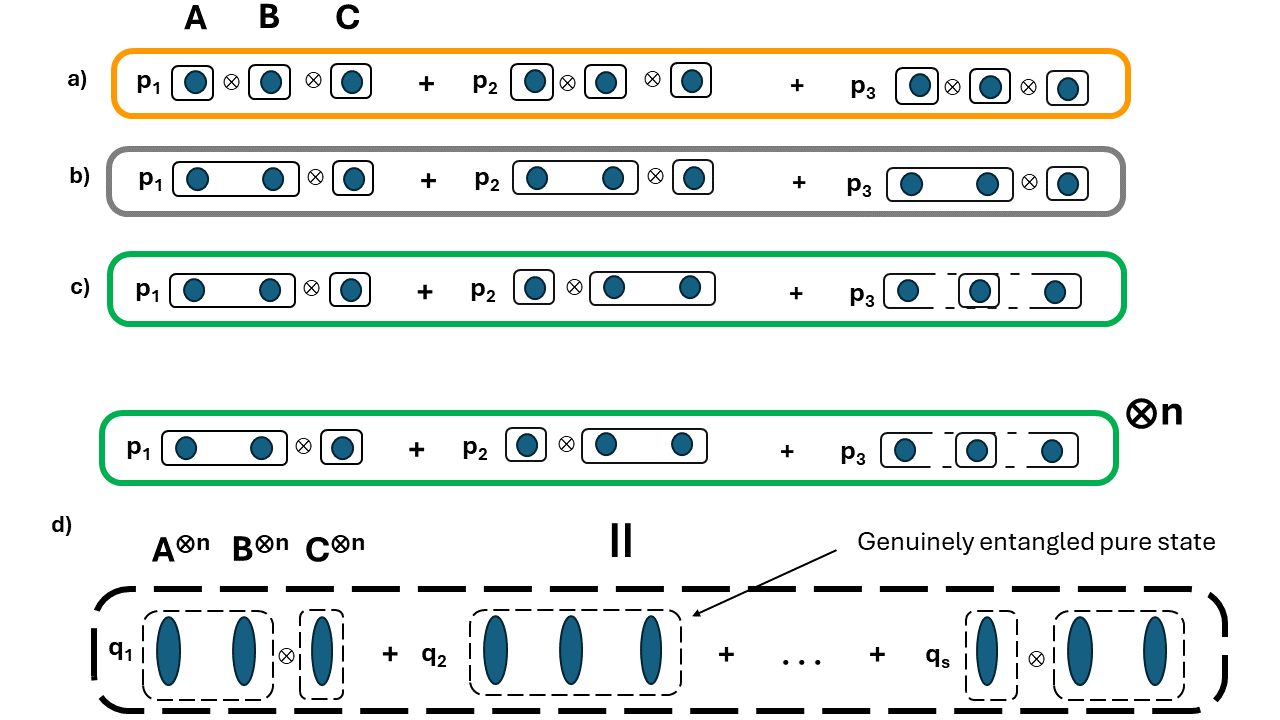} \ \  \ \ 
\vskip -0.3cm
\caption{Entanglement classes for $N=3$ parties: a) fully separable class c$_1$);
b)  $\alpha_{K=3}$-separability class c$_2$) with respect to  partition $S_1={A,B}$ and $S_2={C}$;
c) $K=2$-separability class c$_4$) -- colours stress that the partitions in components of the convex combinations are different.
d) illustration of the fact that  $K$-separability is {\it not tensor-stable}, since it can be {\it activated to a genuine $N$-partite entanglement} \cite{Yamasaki-Et-Al-22,Palazuelos-de-Vincente-22}: if only it is not  $\alpha_{K=2}$-separable 
with respect to any  bipartition 
 then  there exists a number of copies $n$ such that any convex decomposition of the resulting state $\rho^{\otimes n}$ contains {\it at least} one state that is not separable under any cut (marked in red),  equivalently -  the resulting state is not $K=2$-separable 
 and as such it is genuinely entangled (class c$_5$). This effect has no analogue in 
 the bipartite case. 
There are special states (for 3-qubit see $\rho_{UPB}$ \cite{sepEXT2,UPB2}, in Sec. III.D),  which are $\alpha_{K=2}$-separable 
with respect to {\it all} bipartitions, yet they are {\it not fully separable} (see also Fig. \ref{biseparability}). 
The states in b) and c) are $M=2$-producible, and the final state in d) is not $M=2$-producible c$_6$), so it is 
genuinely $M+1=3$-entangled c$_{7}$), which agrees with the definition of genuinely $N=3$-partite entangled state c$_{5}$).
The colours mark the minimal sets to which a given state would belong if the systems were qubits in  accordance with the next Fig. 4. In particular, the state from c) is marked by green since it would belong to the green set of BS states,
while the state b) would be a part of 
the gray set $PS_{AB|C}$. }
      \label{fig_3ab2}
     \end{center} 
 \end{figure}

 \begin{figure} [htbp]
   \begin{center} 
 \includegraphics[width=13cm,angle=0]{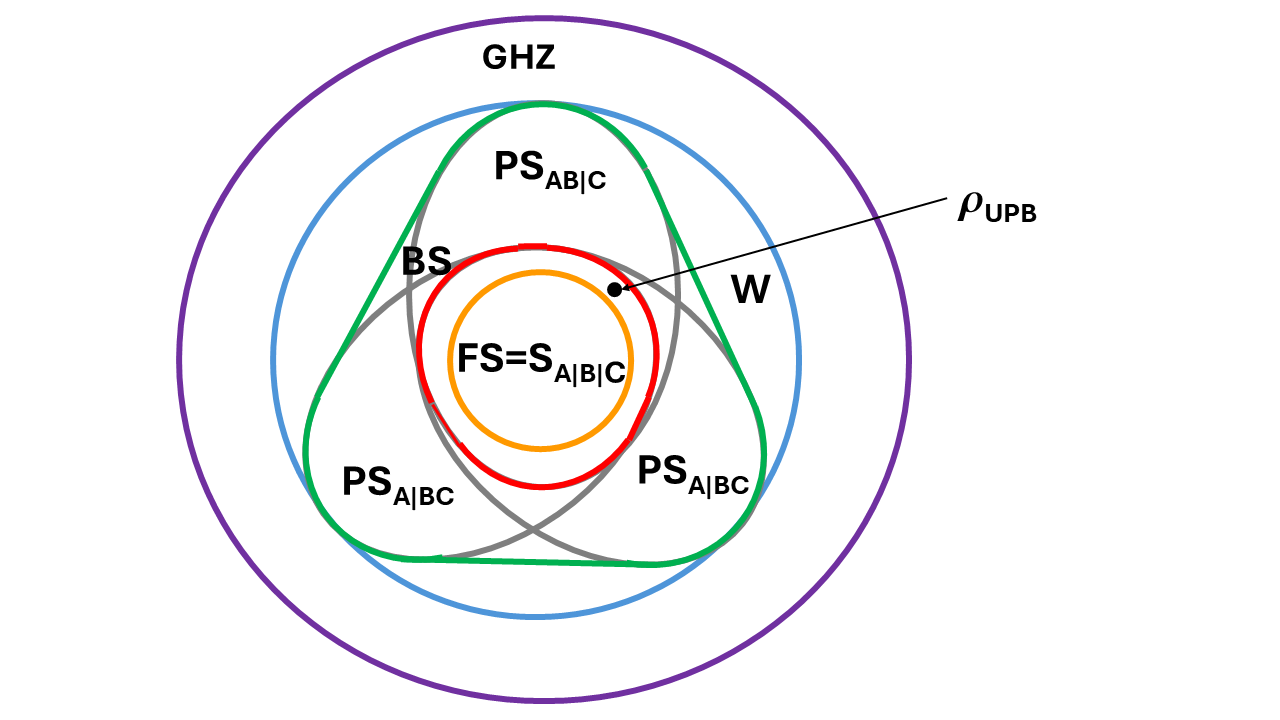} \ \  \ \ 
\vskip -0.3cm
\caption{Four classes of  {\it mixed} 3-qubit states forms convex sets:
$FS \subseteq BS \subseteq W \subseteq GHZ $ (fully separable, biseparable, W and GHZ)
represented by orange, green, blue and purple 
curves respectively.
The biseparable set $BS$ is the convex hull of the 
three different partially separable (PS) classes marked in gray. 
The intersection (red) of the classes is strictly larger than the fully separable (FS) set. Exemplary state $\varrho_{UPB}$, 
discussed in Sec. III D, illustrating this difference is also symbolically depicted.}

      \label{biseparability}
     \end{center} 
 \end{figure}

 Interestingly, there exist a full classification of mixed $3$-qubit states. It refers to the classes of pure states: (i) fully separable class of pure states,  (ii) partially separable class pure states i.e. separable with respect to specific cuts (A|BC, AB|C or B|AC) and - finally - (iii) $W$ and (iv) $GHZ$ classes of pure states defined as a SLOCC orbits with respect to the special linear group (see sec. III.B) of the $| W\rangle$ and $| GHZ\rangle$ states (\ref{GHZ_W}) respectively.
 The corresponding classes of mixed states are (\cite{ABLS01}, see also \cite{Guehne-Toth-2009}): (I) {\it fully separable} (FS) - convex hull of elements from (i) (II) {\it partially separable (PS)  with respect to specific cut} - convex hull of those elements from (ii) that are separable with respect to that cut (III) biseparable (BS) - convex hull of PS with respect to all cuts (IV) mixed $W$ class - convex hull of BS states and elements from (iii)  (V) $GHZ$ class - set of all states equal to a convex hull of $W$ and elements of (iv). 
 The sets obey inclusions,
 $FS \subseteq BS \subseteq W \subseteq GHZ$, which are strict. 
 The 
 intersection of all PS sets is strictly larger than the set FS -- see Fig. 4.

 The problem of deciding whether a given multipartite state belongs to one of the above classes is in general hard -- see \cite{HHHH09,Guehne-Toth-2009,Nature-Phys-Review}. The positive partial transpose (PPT) test \cite{Peres96},
necessary and sufficient for separability of bipartite 
qubit-qubit and qubit-qutrit case \cite{HHH96}), is in general only necessary 
for separability of mixed states with respect to any division into two parties or - equivalently - to  $\alpha_K$-separability for $\alpha_K=\left(S_1,S_2\right)$ (cf. the class  $c_2$ above). However, it is necessary and sufficient for all multipartite pure states. The
PPT test implies that if bipartite state $\varrho_{XY}$ is separable then the matrix
$\varrho_{XY}^{T_{Y}}$ (called {\it partial transposition of $\varrho_{XY}$}) with matrix elements 
$\langle e_m| \langle f_{\mu}| \varrho_{XY}^{T_{Y}} |e_n \rangle |f_{\nu}
\rangle \equiv \langle e_m| \langle f_{\nu}| \varrho_{XY} |e_n \rangle |f_{\mu}
\rangle $ defined for  orthonormal bases $\{ |e_m \rangle  \}$, $\{ | f_\mu  \rangle \}$ has nonnegative 
spectrum, and hence forms a quantum state, too. 

\subsection{Important classes of transformations} 
The key paradigm to practically quantify entanglement, so called local laboratories paradigm, and operations associated with it are: local operations (LO) plus classical communications (CC), called LOCC \cite{BDSW96,CLMOW14}. 
We shall briefly recall the formal construction of LOCC after \cite{CLMOW14}. Let ${\cal L}({\cal B}({\cal H}))$ be a family of all bounded linear maps  (with possibly different co-domains) on a set ${\cal B}({\cal H})$ of all bounded linear maps  on a Hilbert space $\cal H$.  {\it A quantum instrument} ${\cal I}= (\{ {\cal E}_j \}, j \in \Theta )$, $\Theta$ being finite or countable, consists of a family of trace preserving (TP) maps ${\cal E}_j \in {\cal L}({\cal B}({\cal H}))$ such that $\sum_j {\cal E}_j $ is trace preserving (TP). 
The set of instruments is equipped with a metric induced by a diamond norm (see \cite{CLMOW14} for details). Now, given 
a quantum system $\mathcal{H}_{A_1 ... A_N}:= \bigotimes_{j=1 }^{N} \mathcal{H}_{A_j}$ we call an instrument ${\cal I}^{K}=\{ {\cal F}_j  \}$ {\it 1-way local with respect to party K}
iff 
 ${\cal F}_j ={\cal E}_j^{(K)} \otimes \left( \bigotimes_{J\neq K} {\cal T}_{j}^{(J)} \right)$
for some completely positive trace preserving (CPTP) maps ${\cal T}_{j}^{(J)}$. Here, the $K$-th party had applied an instrument $\{ {\cal E}_j^{(K)} \}$ followed by broadcasting the classical outcome $j$ to all the other parties who subsequently applied their CPTP operations $ {\cal T}_{j}^{(J)}$. 
There is a method to compose 1-way local instruments called {\it linking} \cite{CLMOW14}, which allows one to build an arbitrary LOCC instrument (term "instrument" is interchangeable with "operation"). As a result, the list of classes of important operations includes:
(i) LOCC$_0$: zero-way operations $\bigotimes_{J=1}^{N} {\cal E}_{j_{J}}^{(J)}$  where each 
party $J$  (for $J=1, ..., N$; $j_{J}=1, ..., \Theta_J$)  performs an instrument independently;
(ii) LOCC$_1$: 1-way local operations with respect to some party K; (iii) LOCC$_r$ ($r \geq 2$): all operations LOCC linked to some ${\cal I} \in$ LOCC$_{r-1}$;  (iv) LOCC$_\mathbb{N}=
\bigcup_{r \in \mathbb{N}}$ LOCC$_r$; 
(v) LOCC: set of all operations ${\cal I}$ being a limit (in the diamond norm) of some sequence of coarse-grainings ${\cal I'}_{n}$ of ${\cal I}_{n} \in $LOCC$_\mathbb{N}$ where ${\cal I}_{n}$ is LOCC linked with  ${\cal I}_{n-1}$;
(vi)  $\overline{\textrm{LOCC}}$: completion of LOCC in the diamond norm; (vi-vii) separable "SEP" (resp. positive partial transpose "PPT")  --- i.e. {\it separable instruments} (resp. {\it PPT instruments}) preserving separability (resp. PPT) after extension with identity map $\mathbb{I}$ on ${\cal B}(\mathcal{H}_{A_1' ... A_N'})$ which means that the N-party state 
 $\tilde{\rho}_{\tilde{A}_1, ..., \tilde{A}_N}
:=\mathbb{I}_{A_1', ..., A_N'} \otimes {\cal E}_j (\rho_{A_1' A_1:A_2' A_2: ..., A_N' A_N})$
 (defined on the space $\mathcal{H}_{\tilde{A}_1 ... \tilde{A}_N}$ with $\tilde{A}_i$ having an internal structure $A_i' A_i$) is fully separable (resp. PPT with respect to all bipartite partitions).  
The sets of quantum instruments (i-vii) constitute a sequence of sets all with {\it strict inclusions} from the smallest set LOCC$_0$ to the largest set PPT (see \cite{CLMOW14} and references therein).  With the exception of PPT, they all preserve separability classes of states defined in the previous section.
One also frequently considers the coarse-grained {\it LOCC  transformation} $\Lambda_{LOCC}=\sum_j {\cal E}_j$ associated with the LOCC instrument ${\cal I}$.

{\it Remarks: other classes of operations.---} Removing the ,,primed"  copies in  the state $\tilde{\rho}_{\tilde{A}_1, ..., \tilde{A}_N}$ defining SEP and PPT operations above
 leads to the strictly smaller  separability preserving (SEPP) and PPT preserving (PPTP) classes as illustrated by unitary SWAP operation ${\cal E}_{SWAP_{A_i A_j}}$ swapping ${\cal H}_{A_i}$ with ${\cal H}_{A_j}$ (see \cite{Ishizaka-Plenio05} for multipartite applications and \cite{Regula-et-al19} for other PPT related operations). Another class of local operations and shared  randomness (LOSR)  \cite{Schmidt-et-al-24} has applications to Bell inequalities.  Unlike LOCC, these operations  cannot create, so called,  graph states from bipartite entanglement shared among network nodes \cite{Makuta-et-al-23}.

For Hilbert spaces of finite dimensions considered here the Carath{\'e}odory theorem implies that elements of SEP are coarse grainings of the canonical form 
\begin{equation}
{\cal E}_j (\cdot) = \bigotimes_{i} M_{A_i}^{j} (\cdot) (M_{A_i}^{j} )^{\dagger} \ \  \ \text{with}  \ \ \sum_{j \in \Theta} \bigotimes_{i} M_{A_i}^{\dagger} M_{A_i} =  \mathbb{I}_{A_1, ..., A_N}, \ \ |\Theta| \leq dim ({\cal H}_{A_1 ... A_N})^{4}
\label{SEP-not-limit}
\end{equation}
which sometimes  
cannot be physically realised by LOCC  \cite{BDF+99}. 
Furthermore,  an instrument ${\cal I}= (\{ {\cal E}_j \}, j \in \Theta )$ can be performed by {\it Stochastic LOCC 
(SLOCC)} if there exists some probability $p$ such that 
${\cal I}(p)= (\{ {p\cal E}_j \} \cup {(1-p)\cal D}, j \in \Theta )$ is LOCC, with {\it the depolarising map} generally defined on
 $B({\cal H})$ as ${\cal D}(\cdot) :=Tr( \cdot)\frac{Id}{\textrm{dim} {\cal H}} $ and the identity $Id$ on  ${\cal H}$). Any SEP operation with $|\Theta|$ outcomes can be performed by SLOCC \cite{DVC00} with the probability  \cite{CLMOW14} at least 
  $[|\Theta| (\textrm{dim} ({\cal H}_{A_1 ... A_N})]^{-2}$.
There is a crucial observation that the SEP class  as well as its LOCC subclasses {\it preserve all the separability classes of N-partite systems} defined in the previous chapter.

{\it Local unitary operations (LU) as an example of LOCC$_{0}$ and associated symmetries.--- } The transformation $\rho_{A_1 ... A_N} \rightarrow (\otimes_i U_i)\rho_{A_1 ... A_N} (\otimes U_i^{\dagger})$ with unitary operations $U_i$ is an elementary zero-way transformation. Following  \cite{SWGK-PRX-20} one defines 
the composed groups $\tilde{K}=SU(d_1) \otimes ... \otimes SU(d_N)$ and $K=U(d_1) \otimes ... \otimes U(d_N)$ of 
special unitary and unitary matrices 
respectively. Given any subset $H$ of the two groups $H \subset \tilde{K} \subset K$  for any N-partite pure states $\psi$  we define {\it the stabiliser of $\psi$ with respect to $H$} as 
$H_{\psi}=\{ h \in H: h|\psi \rangle  = |\psi \rangle \}$. In two-party case with $d_1=d_2=d$, $ K=U(d) \otimes U(d)$  
the paradigmatic example is the stabiliser $H_{\psi^{+,d}}=U \otimes U^{*}$ ($U \in U(d) $ ) of the maximally entangled Bell state $|\psi^{+,d}\rangle $.
The latter is an example of a critical state, since it has maximally mixed reduced density matrices $\rho_{A_1}=\rho_{A_2}=I_d/d$. The {\it set of critical states $Crit({\cal H}_{A_1 ... A_N})$}, also  called {\it locally maximally entangled (LME) states}, is defined as  containing pure  states with all single-system reduced density matrices being maximally mixed  \cite{Gour-Wallach-NJP-11}.  
N-partite critical state with local dimensions $\{ d_i \}$ exist 
if $R(d_1, ..., d_N):=\Pi_{i=1}^{N}d_i - \sum_{l=1}^{N}(-1)^{l+1} 
\sum_{0 \leq i_1 < ... < i_l \leq N} gcd(d_1^{2},\ldots, d_N^{2}) \geq 0 $~\cite{BLRR-Quantum-19}.

The three-qubit GHZ has a stabiliser is an example of critical state and its stabiliser is $H=\{ \sigma_{x} \otimes\sigma_{x} \otimes \sigma_{x} , 
e^{i\phi_1 \sigma_z} \otimes e^{i\phi_2 \sigma_z} \otimes e^{-i(\phi_1 + \phi_2) \sigma_z} \}, \phi_1, \phi_2 \in [0, 2\pi)$.  
For N-qubit 
case, $N\geq 5$ \cite{Gour-Wallach-NJP-11} and for $N$-qudit states for $N\geq 4$ 
there exists \cite{SWGK-PRX-20} 
 a {\it LME  pure state} $\psi$ with trivial LU stabiliser,
   $\tilde{K}_{\psi}=Id_{A_1}\otimes ... \otimes Id_{A_N}$.
The simplest example for $d=2$ and $N$=5 is 
 $   |\Psi_{2,5}\rangle=\sqrt{7}|00000\rangle +  \mathrm{SYM}(|00111\rangle) + \sqrt{5}|11111\rangle$
where $\mathrm{SYM}(|00111\rangle)$ is symmetrization of the input state,
$\mathrm{SYM}(|001\rangle)=|001\rangle + |010\rangle + |001\rangle$.

{\it Quantum teleportation (and entanglement swapping) as an element from LOCC$_1$.---} 
This is bipartite LOCC acting on a the bipartite system 
$\tilde{A} B := A' A B$ where the composite  $ d \otimes d$
subsystem $\tilde{A}:= A' A$ is under control of the 
experimentalist called Alice and the d-dimensional system B is 
controlled by Bob. There is a 1-way LOCC instrument ${\cal F}_{mn}^{{\tilde A}B} = {\cal P}_{mn}^{{\tilde A}} \otimes  {\cal U}_{mn}^{B}$, $j \in \Theta = {1, ..., d^{2}}$, $j=(mn)$, $m, n \in 0, ..., d-1$ with ${\cal P}_{mn}$ being an Alice measurement projecting on the  maximally entangled basis $\{ |\Psi_{mn} \rangle = [U_{mn}^{A'} \otimes Id^{A} ] |\psi^{+,d} \rangle_{A'A} \}$, with unitaries $U_{mn}=\sum_{k=1}^{d} e^{2 \pi i k n / N} |k \rangle \langle (k + m) \textrm{mod} N $. Here Bob performs unitary rotation $ {\cal U}_{mn} (\cdot)=
U_{mn}(\cdot)U_{mn}^{\dagger}$. The above 1-LOCC operation followed by tracing out Alice systems, if applied
to the state $|\phi_{A'}\rangle |\psi^{+,d}_{AB} \rangle$,  transfers the state 
$|\phi \rangle$ to Bob's lab \cite{Bennett-teleportation-93}, the procedure which is called {\it quantum teleportation}. 

By linearity one may  extend ${\cal F}_{mn}^{{\tilde A}B}$ to $\tilde{\cal F}_{mn}^{X{\tilde A}B}=\mathbb{I}^{X}\otimes {\cal F}_{mn}^{{\tilde A}B}$ 
with some third lab denoted by $X$ and then  the same protocol maps 
any state of the form $\rho_{XA'} \otimes \psi^{+,d}_{AB}$ into $\rho_{XB}$ effectively {\it swapping} the part $A'$ into Bob's lab. If the state $\rho_{XB}$ is entangled we call 
the protocol {\it entanglement swapping}. This protocol entangles the particle $X$ with Bob's particle $B$, 
even though they have never interacted before -- see \cite{Zukowski}.

{\it Local filtering of a quantum state as the simplest SLOCC transformation.--- }
A tool, mentioned already in previous sections and exploited especially in analysis of multipartite pure entanglement, is {\it local filtering}. It can be introduced by considering the following two-element instrument ${\cal I}_{filtering}=\{ {\cal E}_1 ,   {\cal E}_2 \}$
\begin{equation}
 {\cal E}_1 (\cdot) = \otimes_{i} L_{A_i} (\cdot) (L_{A_i})^{\dagger} ,   \ 
 {\cal E}_2 (\cdot) = Tr( (Id_{A_1 ...A_N} -\otimes_{i} L_{A_i}^{\dagger} L_{A_i}) 
 (  \cdot  ) ) \frac{Id_{A_1 ...A_N}}{  {\textrm{dim}} {\cal H}_{A_1 ... A_N}   }
 \ \ \text{with} \ {L_i}^{\dagger} L_{i} \leq I_{dim {\cal H}_{A_i}}
\label{local-filtering}
\end{equation}
where the latter condition satisfied for all $i=1, ..., N$ makes the whole transformation TP.
The above can be easily reproduced in LOCC protocol as a sequence of 1-way LOCC 
operations \cite{CLMOW14}.

Finally, {\it asymptotic conversion (respectively - production)} of a given state $\rho$ into the 
states $\sigma_1, ..., \sigma_l$ {\it with respect to a class ${\cal C}$} of maps (usually LOCC) achieves {\it conversion (respectively - production) rates $r_1, ..., r_l$} if
there is  $\Lambda^{(n)} \in {\cal C}$ such that
\begin{eqnarray}
lim_{n \rightarrow \infty} 
||
\Lambda^{(n)} (\rho^{\otimes n})- \sigma_1^{\otimes \lfloor r_1 n \rfloor } \otimes ... \otimes \sigma_l^{\otimes \lfloor r_l n \rfloor }||_1
=0, 
\label{Conversion}
\end{eqnarray}
and respectively
\begin{eqnarray}
 lim_{n \rightarrow \infty} 
 ||\Lambda^{(n)}(\sigma_1^{\otimes \lfloor r_1 n \rfloor } \otimes ... \otimes \sigma_l^{\otimes \lfloor r_l n \rfloor })- \rho^{\otimes n}||_1
 =0.
 \label{Production}
\end{eqnarray}
Usually, the states $\{ \sigma_{i} \}$ represent well-defined resourceful states like some maximally entangled states of different types. The following surprising  example is extremely instructive.

{\it Quantum state merging and quantum combing.--- } Let us consider an asymptotic conversion of a state $|\Psi \rangle_{ABC}$  (where A, B, C correspond to different labs) into the same state $|\Psi \rangle_{B'BC}$ but now with the subsystem A transferred to Bob's lab by  LOCC  executed only by Alice and Bob.
One may just exploit {\it swapping} which consumes $r_1=1$ pair of $\boldsymbol{\psi^{+,d}}_{A'B'}:=\sigma_1$ per swapping of a single system $A$ to the Bob's lab (here and below we use the convention $|\psi\rangle \langle \psi|= \boldsymbol{\psi}$).

An asymptotic protocol called {\it quantum state merging} \cite{State-merging}  consumes typically much less entanglement, namely only  $r_1=S(A|B)$ of
pairs $\boldsymbol{\psi^{+,d}}_{A'B'}:=\sigma_1$  
together with $r_2=1$ of the original
$ \boldsymbol{\Psi}_{ABC}:=\sigma_{2}$ to produce 
$\boldsymbol{ \Psi}_{B'BC}:=\rho$ in terms of asymptotic production (\ref{Production}).
The quantum conditional entropy $S(A | B)=S(\rho_{AB}) - S(\rho_B)$,
expressed in terms of von Neumann entropies of 1- and 2-partite marginals of the total state $\boldsymbol{\Psi}_{ABC}$, 
is a {\it quantum analog of classical missing information} about A conditioned upon known B. This clear analogy, however, breaks down when $S(A|B)<0$  since missing information cannot be negative. Surprisingly, in this case the second variant of quantum state merging \cite{State-merging}  converts, in the sense of (\ref{Conversion}), the input state $\rho=\boldsymbol{\Psi}_{ABC}$ to the output states 
 $\sigma_1=\boldsymbol{\Psi}_{B'BC}$, $\sigma_2=\boldsymbol{\psi^{+,2}}_{A'B'}$, 
with $r_1=1$ and {\it an entanglement production rate} $r_2= - S(A | B) >0 $ (in terms of e-bits)
(for multipartite variant and the corresponding {\it regions  of admissible rates} see \cite{Multiparty-state-merging}). The above was extended to {\it entanglement combing} \cite{Combing-Yang-Eisert}, where the state $|\Psi \rangle_{A, B_1, ..., B_N}$ was asymptotically LOCC-converted with rate $r=1$ into $|\phi_1 \rangle_{A_1 B_1} \otimes  ...
|\phi_N \rangle_{A_N B_N}$ with $\sum_{k=1}^{N} S(\rho_{A_{k}})= S(\rho_{A})$. 

\subsection{Mathematics of filtering SLOCC transformations of pure states}
The set of single-Kraus-operator SLOCC or {\it filtering} SLOCC in the case of pure states' transformations is usually referred to just as SLOCC. Analogous to local unitary 
$\tilde{K}$ and local special unitary $K$ transformations one defines 
$\tilde{G}=GL(d_1) \otimes ... \otimes GL(d_N)$ and $G=SL(d_1) \otimes ... \otimes SL(d_N)$ with 
$GL(d)$,  $SL(d_N)$ standing for general linear matrices and matrices from special linear group on $\mathbb{C}^{d}$ respectively, as well as the corresponding stabilisers $G_{\psi}$ and $\tilde{G}_{\psi}$. 
The original filtering SLOCC transformation 
maps an $N$-party pure vector $|\psi_{A_1 \cdots A_N} \rangle $ as
\begin{equation}
\psi_{A_1 \cdots A_N}  \rightarrow \psi^{'}_{A_1 \cdots A_N} =\frac{\otimes _i L_i |\psi_{A_1 \cdots A_N} \rangle }{|| \otimes _i L_i |\psi_{A_1 \cdots A_N} \rangle  ||}
\label{pureSLOCC}
\end{equation}
with the probability $P(\psi_{A_1 \cdots A_N} \rightarrow \psi^{'}_{A_1 \cdots A_N})= 
|| \otimes _i L_i |\psi_{A_1 \cdots A_N} \rangle  ||^{2}$ and  $\otimes _i L_{i} \in \tilde{G}$ (i.e.  $L_{i}$ may be not reversible) and satisfying 
[c.f (\ref{SEP-not-limit})] $L_i^{\dagger}L_{i} \leq Id_{A_i}$. 

An important quantity characterising any pure state and related to the above is called {\it  Schmidt rank} in bipartite case and {\it tensor rank} or generalised Schmidt rank in the multipartite case; defined as  $R_{S}(\psi)$, it is the minimal number of product states in such a decomposition and it can be extended to mixed states by {\it convex roofs technique} -- see \cite{HHHH09}.  
For bipartite pure states it equals the rank of either of the reduced states and can be easily calculated. 
 For a larger number of subsystem this calculation becomes NP-hard \cite{Hastad-90}. However, it is {\it a SLOCC monotone} i.e. it does not increase under the above general SLOCC transformation (\ref{pureSLOCC}) \cite{Lo-Popescu-01}, being preserved if 
 $\otimes_i L_i \in G$ (i.e. if all $L_{i}$ are reversible). In analogy to entanglement polytopes, 
 the vectors of Schmidt ranks have been defined both for pure as well as mixed multipartite states featuring non-trivial constraints  -- see \cite{Huber-DeVincente-2013} and references therein.
 
Below, we consider SLOCC with respect to local group $G$ (which have a determinant one). Generally, we call the set $G|\psi \rangle= \{ \phi: \phi=g \psi, g \in G \}$ a G-orbit of vector $\psi$. Two vectors $\psi$ and $\phi$ are {\it SLOCC convertible (with finite probability)} if their $G$-orbits are the same, i.e. when $g\in G$ exists such that $|\psi \rangle = g |\phi \rangle$ or $g^{-1}|\psi \rangle = |\phi \rangle$.  Clearly, since $g \in G$ is invertible, both $g=\otimes _i L_i $ and $g^{-1}=\otimes _i L_i^{-1}$ can be rescaled to satisfy $L_i^{\dagger}L_{i} \leq Id_{A_i}$ and define physical SLOCC. For permutational symmetric pure states the symmetric SLOCC operations are sufficient \cite{Migdal}.

{\it Genuinely (or truly) N-partite entangled pure states}
are all pure states which are bipartite entangled under any cut. 
Consider now only pure states with local reduced density matrices
of maximal rank. There is only one orbit defining them in the 
case of any bipartite $d \otimes d$ case, namely, $G|\psi^{+,d}\rangle$. 
For three-qubit systems there are two orbits corresponding to $G|GHZ \rangle $ and $G|W \rangle$ states, which means that genuinely entangled states can be entangled in two inequivalent ways \cite{DVC00}. In case of 4 gubits there are
already infinitely many inequivalent G-orbits, divided into 9 inequivalent classes \cite{VDDV02}.
For  instance, states
$|\phi(a) \rangle:=a(|0000\rangle + |1111\rangle) + (|0011\rangle + |0101\rangle + |0110\rangle$
with different $a$ are not SLOCC interconvertible, or equivalently, 
$G|\phi_a \rangle$ differ for 
different complex numbers $a$.
For classification of
SLOCC orbits for
$2 \otimes m \otimes n$  systems
with help of {\it $m \times n$ matrix pencils} see \cite{SlowikHKS-Quantum-20}.

Interestingly, {\it  locally maximally entangled (LME) states} (also called {\it critical states}, see section III.A), defined by their maximally mixed local density matrices, are special. 

There is a {\it unique LU orbit of vectors with a norm $1$} within SLOCC orbit of any of them and they serve as a representative of this LU orbit (this follows from the Kempf-Ness theorem, see \cite{Monograph}). Furthermore, 
the union of their orbits, which
consists of normalised states $  (G | \psi \rangle)^{norm} =\{ |\phi \rangle : |\phi  \rangle = \frac{g |\psi \rangle}{ ||g |\psi \rangle||}\;\; \mathrm{s.t.}\;\; g \in  G \ \text{and} \  |\psi \rangle \in Crit({\cal H}_{A_1 ... A_N}) \}$, is {\it dense} in the projective space $\C P^{d-1}$, $d=\Pi_{i=1}^{N}d_i$.
 In the case of 
$N >3$ qubits there are infinitely many LU orbits of LME states parametrised by $2^{N} - 3N - 1$ parameters, while for $3$ qubis there exist only one - that of the state $|GHZ\rangle$. Hence, up to local unitary transformations, this is the unique state approximating  {\it  all} $3$-qubit pure states (including $|W \rangle$) in the above sense. This shows a salient difference between bipartite and multipartite states: the state $|GHZ\rangle$ has smaller  multipartite Schmidt rank than $|W\rangle$, yet it is more 'effective'. 
Critical states have distinguished local spectra: all of them are uniform. In general, the local spectra  $\vec{\lambda}_{1}, ..., \vec{\lambda}_{N}$
of general  N-partite pure states exhibit the polytope structure which reflects the partial 
order of SLOCC orbits in terms of inclusions of the corresponding {\it entanglement polytopes} -- see \cite{Klyachko04,Sawicki1,WDGC13,Sawicki2} (for asymptotic properties of qubit entanglement polytopes see \cite{Sawicki3}).
 
\subsection{Deterministic LOCC transformations of pure states and the issue of maximally entangled states}

There is some ambiguity in defining maximal entanglement in multipartite case. The most natural definition of {\it the set of maximally entangled vectors} in ${\cal H}_{A_{1}...A_{N}}$ inherited after bipartite case seems to be clear: those are the vectors (up to LU equivalence) such that any other vector from  ${\cal H}_{A_{1}...A_{N}}$ can be produced from them deterministically by LOCC associated with the {\it original space} ${\cal B}({\cal H}_{A_{1}...A_{N}})$ as an input. 
Hence, Nielsen majorisation \cite{Ni99} guarantees that bipartite maximally entangled states correspond to a single LU, $U(d_{1}) \otimes U(d_{2})$ orbit $\tilde{K}|\psi^{+,d}\rangle$ with maximal Schmidt rank 'reference  state' $|\psi^{+,d}\rangle$  ($d=\text{min} \{ d_1, d_2 \}$). For three qubits the $3$-parameter family has been identified here \cite{Vincente-et-al-13}. 
There are  other  concepts of {\sl multipartie maximal entanglement}. First, 
extending LOCC input by ancilla, the systems with the dimensions satisfying 
condition $d_{1}\geq \Pi_{k=2}^{N}d_{i}$
have a single (up to LU) maximally entangled state 
of a combing-like structure $|\Psi \rangle =  \otimes_{i=2}^{N}|\Psi^{+,d_i}\rangle_{A_{1,i} A_{i}}$. Indeed, it  can serve to teleport $N-1$ parts of {\it any} $N$-partite 
pure ancilla state from the lab number $1$. 
Another variant of  maximal entanglement in a weaker sense are  {\it locally maximally entangled} (LME) states, 
as  they are SLOCC dense in the projective space. 
A unique element among them, frequently considered as maximally entangled N-partite state, is the GHZ, since LOCC can deterministically produce bipartite maximally entanglement  ($|\psi^{+,d} \rangle$) between {\it any} two nodes out of it \cite{BPRST00}. Finally, a stronger variant of  LME are {\it absolutely maximally entangled states} (AME) which have all the reductions up to $\lfloor \frac{N}{2} \rfloor$ maximally mixed
\cite{Sc04,FFPP08}. 
Hence, for any bipartite splitting both 
parties are maximally entangled 
\cite{HCLRL12,AC13,GALR15}. 
In the case of $N$ subsystems with $d$ levels each,
the following conditions are necessary for AME to exist:
$N\leq 2(d^{2}-1)$ (N-even) and $N\leq 2(d (d+1)-1$ (N-odd).
There are neither AME states for $N=4$ qubit system \cite{HS00}, nor for $N\ge 7$ qubits \cite{HGS17,Huber-et-al-18}. AME states do exist 
for $N=4$ systems with $d\ge 7$ levels
and if  $N\leq d$ and $d$ equals a power of a prime number \cite{Grassl-et-al-04}. A recently
discovered AME state for four subsystems
with $d=6$ levels is related to the solution
of the quantum analogue of the problem of 
$36$ officers of Euler \cite{R+22}.
In several cases the existence of AME states remains open. A current list
of known constructions is available online \cite{HW_table}.

Let us mention that all AME states of qubits are some special instances of so called graph states of 2, 3, 5 and 6 qubits \cite{HGS17}. Given a graph $G(V,E)$ with vertices $V$ ($|V|=m$) and edges $E$, the corresponding {\it graph state} of $m$ qubits is defined \cite{Raussendorf-2003} by preparation each qubit in the state  $|+\rangle=\frac{1}{\sqrt{2}}(|0\rangle + |1\rangle)$ followed by application of  the controlled-phase gate  $|0\rangle\langle 0| \otimes I +|1\rangle\langle 1| \otimes \sigma_{3}$ to the pairs of qubits corresponding to edges of the graph $G(V,E)$. The set of  graph states of m qubits is described by only $m(m-1)/2$ discrete parameters. The subclass of graph states (called {\it cluster states}) is the crucial resource for one-way quantum computer \cite{Raussendorf-Briegel-2001}.
 
The subset of critical states with maximal G-orbit (i.e. of the same dimensionality as G) and trivial stabilisers, if nonempty, is open, dense and of full measure in ${\cal H}_{A_1 ... A_{N}}$  \cite{Gour-Kraus-Wallach-17}. The non-emptiness was proven for $N>4$ qubits  \cite{Gour-Kraus-Wallach-17},
for $N=3$ with $d=4,5,6$
and for  $N>3$ and arbitrary $d$   \cite{SWGK-PRX-20} leading  a dramatic consequence:
since states having
trivial stabiliser in $G$ are {\it LOCC-isolated} (i.e. are deterministically LOCC interconvertible only within their own LU orbit \cite{Gour-Kraus-Wallach-17}) in all those cases, as opposed to bipartite, a random pure state is useless to deterministically create 
a pure state from a different LU orbit
by LOCC transformations. 
Deterministic pure states  LOCC transformations have their SEP analogs, fully characterised, including the transition probability $P(|\psi\rangle \rightarrow |\phi\rangle )$, 
which were used 
to obtain the above results \cite{Gour-Wallach-NJP-11}.

{\it Entanglement (LOCC) catalysis.---} In the bipartite case, deterministic Nielsen transformation  $|\psi \rangle \rightarrow |\phi \rangle$ works only when the majorisation of the local spectra is fulfilled  $\vec{\lambda}(\psi) \prec \vec{\lambda}(\phi) $. This condition is usually not met, so there are a lot of states that cannot be deterministically LOCC turned one to another. However, 
there is entanglement catalysis phenomenon,
\cite{Jonathan-Plenio-1999} namely for $d \otimes d $
with $d\geq 4$ (and only for them) it is possible 
that $|\psi\rangle \nrightarrow |\phi\rangle$ however $|\psi \rangle \otimes |\eta\rangle \rightarrow  |\phi\rangle \otimes |\eta \rangle$. For qudit GHZ-like states  $| GHZ_{N}^{d} (\vec{\lambda}) \rangle = \sum_{i=0}^{d-1}
\sqrt{\lambda_{i}} |i \rangle ^{\otimes N}$  the 
direct analog of both Nielsen majorisation  and the original 
catalysis were derived \cite{NGHK21}. 
There exists multipartite  SLOCC catalysis \cite{ChenCDZW10}, a consequence of the fact that 
a  {\it Schmidt rank of multipartite pure states is in general not multiplicative} (unlike for bipartite pure 
states) sharing this feature with bipartite mixed states  \cite{TerhalHorodecki}. In particular,
an  existence of  multicopy SLOCC transformation $|\phi \rangle^{\otimes n} \rightarrow  |\psi \rangle^{\otimes n} $ 
always implies possibility of catalysis. 
\subsection{Asymptotic transformations}

\begin{figure} [htbp]
   \begin{center} 
 \includegraphics[width=13cm,angle=0]{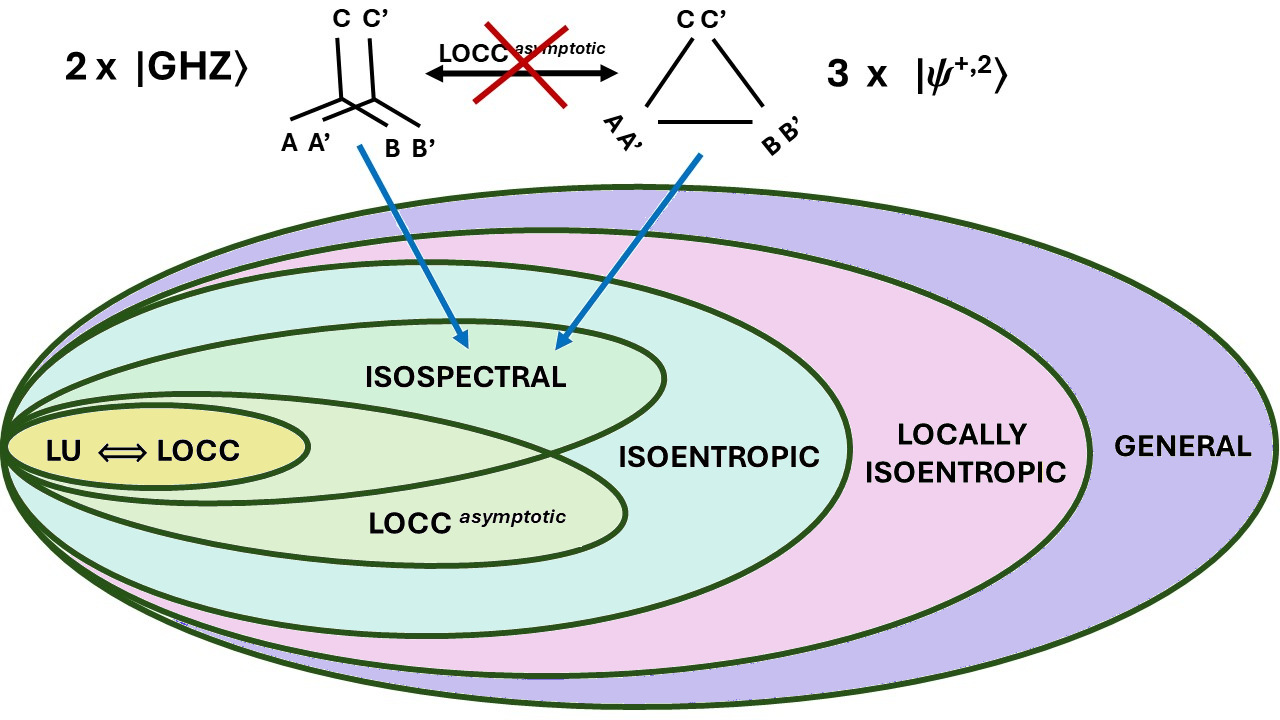} \ \  \ \ 
\vskip -0.3cm
\caption{Pure-states equivalence classes in terms of spectra of their reductions and 
entropies which are relevant for asymptotic LOCC interconvertibility (after Ref. \cite{BennettBPS96}). At the top right corner the pair of 3-partite (in sense of LOCC labs of Alice, Bob and Charlie) states $\{ 2 |GHZ\rangle, 3 |\psi^{+,2} \rangle \}$ which is not interconvertible  despite being not only isoentropic, but even isospectral.
}
      \label{fig_3ab3}
     \end{center} 
 \end{figure}

As already mentioned in the bipartite case, many states cannot be deterministically 
LOCC interconverted. This problem can be removed in the asymptotic regime 
since any entangled pure state $\boldsymbol{\psi}$ can be {\it asymptotically reversibly} 
LOCC-transformed into $\boldsymbol{\psi^{+,2}}$ with the rate 
$r=E(\psi)=S(\rho^{A}_{\psi})=S(\rho^{A}_{\psi})$; This protocol is called 
{\it entanglement concentration} \cite{BennettBPS96}. Remarkably,  only a single function $E(\psi)$ of local spectra plays a (crucial) role here. Summarising, maximally entangled 2-qubit  (or, equivalently, changing the log base - $2$-qudit) state is a unique unit of bipartite entanglement. The situation is much more complicated in the case of multipartite entanglement
\cite{BPRST00}. In this case one needs to compare 
(i) spectra  and (ii) entropies of all the reductions,  including 
(iii) entropies of minimal subsystems --- called local entropies. 
Pure states having the same (i), (ii), (iii) are called {\it isospectral}, 
{\it isoentropic}  and {\it locally isoentropic}. 
While strictly LOCC interconvertible states are only those 
that are LU equivalent, some  {\it isospectral} ones (see Fig. \ref{fig_3ab3}), namely, 
$|GHZ \rangle_{ABC} |GHZ \rangle_{A'B'C'}$ and $| \psi^{+,2}  \rangle_{A'B} |\psi^{+,2}  \rangle_{B'C} 
| \psi^{+,2}  \rangle_{C'A} $  are, surprisingly, neither strictly \cite{BPRST00} nor even asymptotically \cite{Linden-et-al-05} LOCC  interconvertible! 
Hence the question arises: what is  {\it the  minimal reversible entanglement generating set} (MREGS) -- the minimal set of 
entangled "units" into which one can reversibly convert any other 
entangled pure state? In the bipartite case $|\psi^{2,+} \rangle$ there is a single element MREGS, since  any state \cite{BennettBPS96} $|   \psi_{AB} \rangle$ satisfies  both  (i) 
$lim_{n \rightarrow \infty} \left( \inf_{\Lambda \in {\cal C}}  ||
\Lambda( \boldsymbol{\psi}_{AB}^{ \otimes n})- (\boldsymbol{\psi^{2,+}})^{\otimes \lfloor E(|{\psi}_{AB}\rangle) n \rfloor } ||_1\right)=0$ as well as (ii)
$lim_{n \rightarrow \infty} \left( 
\inf_{\Lambda \in {\cal C}}  ||
\Lambda(  (\boldsymbol{\psi^{2,+}})^{\otimes \lfloor E(|{\psi}_{AB}\rangle) n \rfloor }) -  \boldsymbol{\psi}_{AB}^{ \otimes n}||_1\right)=0$.

However, even in the case of 3 qubits the MREGS is not known. For example, it is not ${\cal S}=\{ |GHZ \rangle_{ABC}, |\psi^{+,2} \rangle_{AB}, |\psi^{+,2} \rangle_{BC}, |\psi^{+,2} \rangle_{CA} \}$ \cite{Acin-et-al-03} since it cannot generate any $\Phi_{A'B'C'}$ with a PPT reduction $\rho_{A'B'}$, which has {\it the edge property}, i.e. it has no product vector in its range \cite{Bruss-et-al-02}. 

 {\it Transformations of mixed states: entanglement distillation, bound entanglement and superadditivity.---} Filtering SLOCC operation of $\rho$  naturally defines the orbit by normalisation $G(\rho_{A_1 \cdots A_N})^{norm}  = \{ \sigma: \sigma = \rho_{A_1 \cdots A_N} = \frac{\otimes _i L_i \rho_{A_1 \cdots A_N} \otimes_i {L_i}^{\dagger} }{Tr(\otimes _i L_i \rho_{A_1 \cdots A_N} \otimes _i {L_i}^{\dagger})} \}$. For the state $\rho = (1-p) \psi^{+,2} + p|0,1\rangle \langle 0,1|$ ($0 < p < 1$)  the closure  $\overline{G(\rho)^{norm}}$ contains all 2-qubit entangled pure states (phenomenon called {\it entanglement quasi-distillation} \cite{Brassard-Mor-Horodecki-04}, see \cite{Baio-et-al-19} for its  limitations)  reflecting the fact that $\overline{G (|W \rangle \langle W|)^{norm} } $  contains all 2-qubit entangled states (e.g. corresponding to $|\psi^{+,2}\rangle|0\rangle$, etc.) in analogy to GHZ reproducing all 3-qubit states in this way. Similarly, mixture of GHZ with any single excitation state (like $|100\rangle $) can be seen quasi-distillable.

Asymptotically, in the bipartite case {\it entanglement distillation} is a process converting in the sense of (\ref{Conversion}) a given state $\rho$ into a state of maximal entanglement. It is well defined since  there is only one unit of entanglement $|\psi^{+,2}\rangle $ in log base 2 (resp. $|\psi^{+,d}\rangle $ in log base d) units. Bipartite {\it distillable entanglement } is an optimal rate $D(\rho)=\sup \{ r:  lim_{n \rightarrow \infty}  \inf_{\Lambda \in {\cal C}} ||\Lambda( \rho^{ \otimes n})- (\boldsymbol{\psi^{2,+}})^{\otimes \lfloor r n \rfloor } ||_1 =0 \}$ while the bipartite {\it entanglement cost} is  $E_{cost}(\rho)=\inf \{ r:  lim_{n \rightarrow \infty}  \inf_{\Lambda \in {\cal C}} ||\Lambda( (\boldsymbol{\psi^{2,+}})^{ \otimes n})- \rho^{\otimes \lfloor r n \rfloor } ||_1 =0 \}$. 
 Clearly, for bipartite separable states $D=E_{cost}=0$ since LOCC preserves separability and as such, cannot create entanglement out of it. In the case of multipartite states there is an ambiguity of maximal entanglement, so in general one might consider vectors of rates of different mutually LOCC non-equivalent to critical (LME) states (for example states from Fig. 3). Usually, one defines distillability in terms of GHZ states \cite{Maneva-Smolin-02}, stating, that $\rho$ is k-partite distillable iff one can distill k-partite GHZ states among some $A_{i_{1}} ...A_{i_{k}}$ of its N-parties 
 with nonzero {\it GHZ distillable entanglement} $D_{A_{i_{1}} ...A_{i_{k}}}^{GHZ}(\rho)$ defined in full analogy to bipartite $D(\rho)$, replacing
 $|\psi^{+,2}\rangle$ with k-qubit $|GHZ\rangle_{A_{i_{1}} ...A_{i_{k}}}$.
 Distillability to states other than GHZ was also investigated with help of quantum error correction (see \cite{Maneva-Smolin-02,Dur-Briegel-07}).  There is also another option of  {\it random distillation of bipartite entanglement}, investigated for W states \cite{Fortescue-Lo-07,Chitambar11}.
 
 Bipartite PPT entangled states have been discovered to be non-distillable – the phenomenon called bound entanglement \cite{HHH98}, yet it showed remarkable superadditivity effects: a qutrit non-quasidistillable  mixed state becomes quasidistillable if supplied with unbounded amount of PPT entangled pairs \cite{HHH99}. This, and similar phenomena are called {\it activation of bound (nondistillable) entanglement} including  a surprising channel superadditivity \cite{Science}). 
For multipartite case, some general rules are inherited, for instance, it is true that from $\alpha_K$-separable state  with $\alpha_K=\left(S_1,\ldots,S_K\right)$ no k-partite entanglement can be (quasi)distilled with parts belonging to more than one set $S_i$.  The same is true if the state is PPT with respect to all parties of indices  $S_{i}$ (through trivial extension of bipartite case \cite{HHH98}). 
Yet, there are more options and some analogs of  activation effects are surprising. First, there exists unlockable bound entanglement \cite{Smolin01}, namely, the (permutationally symmetric and hence separable under any cut, ergo nondistillable) state $\varrho^{unlock}_{ABCD}=\frac{1}{4}\sum_{i=1}^{4} \boldsymbol{\Psi^{i}}_{AB} \otimes \boldsymbol{\Psi^{i}}_{CD}  $ (with rank-one projectors $\boldsymbol{\Psi^{i}}$ corresponding to the {\it two-qubit Bell basis}: $|\Psi^{1/2}\rangle=\frac{1}{\sqrt{2}}(|00\rangle \pm |11\rangle)$,
$|\Psi^{3/4}\rangle=\frac{1}{\sqrt{2}}(|01\rangle \pm |10\rangle$) can be unlocked, i.e. joining any two states allows deterministic LOCC production of maximal entanglement in the remaining pair. Moreover, tensor product of  five states produced from $\varrho^{unlock}_{ABCD} \otimes \sigma_{E}$ ($\sigma$  - arbitrary) by cyclic permutation of subsystems, becomes fully (i.e. 5-party) distillable --- the effect called superactivation since each of the 5 states was totally nondistillable \cite{Shor-et-al-03}. Finally, there is a class that is not only bipartite PPT but even biseparable under any cut, yet entangled, such as the state $\rho_{UPB}$, which is proportional to the  projector orthogonal to, so called, multipartite {\it unextendible product bases} \cite{sepEXT2,UPB2} $S_{\bf UPB}=\{|0\rangle |0\rangle |0\rangle,|+\rangle|1\rangle|-\rangle,|1\rangle|-\rangle|+\rangle,
|-\rangle|+\rangle|1\rangle \}$, [with $|\pm\rangle=\frac{1}{\sqrt{2}}(|0\rangle\pm
|1\rangle)$]. Note that here biseparability is stable under tensoring, which is not true in more general case (see the end of Sec. II) and the corresponding instability is called activation \cite{Yamasaki-Et-Al-22,Palazuelos-de-Vincente-22} 
since it resembles to some extend superactivation
described above. Note
that passing PPT test under any cut does not need to be a strong requirement: there exist genuine multipartite entangled states with this property \cite{Lancien-et-al-2015}.
However, if in the definition of biseparability one abandons the requirement of all the elements in the convex combination being product, and adopts the PPT property instead, then any 3-qubit permutationally symmetric state turns out biseparable \cite{Novo-2013} (semidefinite programming is a important tool for such relaxation).

\subsection{Measures of multipartite entanglement}

There are well established axioms concerning measures of quantum entanglement \cite{HHHH09}:
(i)  {\it monotonicity under LOCC operations}, 
$E\left(\Lambda_{LOCC}(\rho)\right) \leq E(\rho)$, and (ii) vanishing on separable states, $E(\rho_{sep})=0$ together implying $E(\rho)\geq 0$. Sometimes one requires {\it strong LOCC monotonicity} which is (i') monotonicity on average $\sum_j p_j E(\rho_{j}) \leq E(\rho)$, where $\rho_j=\frac{{\cal E}_j(\rho)}{Tr{\cal E}_j(\rho)}$, $p_j=Tr{\cal E}_j(\rho)$. As for the LOCC class one chooses either fully general LOCC or {\it LOCC with fixed dimensions}, for instance mapping  the
set of $N$-qubit states into itself.

In general, one might choose (ii) on some of the multipartite separability classes c1)-c4) defined in Sec. III, as all of them are preserved by LOCC operations.
There is a concise sufficient
criterion  \cite{MH-05,DD-et-al-06} based on mutually orthogonal pure states 
$\phi_{{\tilde{A}_k}}^{i}$, called {\it local flags}.
 Namely any $f$ 
invariant under LU and 
affine under extension by  local flags (i.e. satisfying $\sum_i p_i f(\rho_i\otimes \phi_{{\tilde{A}_k}}^{i})=\sum_i p_i f(\rho_i)$, with the flags added to $k$-th subsystem $A_{k}$, $k=1,\ldots, N$) is a {\it strong entanglement monotone} as it  satisfies the monotonicity axiom (i').
Two local flags are enough: $i=1,2$ \cite{DD-et-al-06}. 

Bipartite measures naturally inspire analogs of distillable entanglement and distillable cost in terms of $GHZ$ but also other entanglement measures. 
One considers the sum of some chosen bipartite measures, which leads to  {\it global entanglement} \cite{Meyer-Wallach-02}. Other quantities are related to {\it tangle}, $\tau=C^2$,
see Eq. (\ref{tau}), well defined for a fixed dimension. 
The most important technique is 
based on the {\sl convex roof}
construction analyzed by Uhlmann \cite{Uhlmann-roof}: after defining a measure on pure states,  $E(\psi)=f(|\psi\rangle)$,
for some carefully chosen function $f$ one  extends it to mixed states
by 
$E(\varrho)=\inf \sum_ip_i f(|\psi_i\rangle)$,
with infimum taken over all {\it pure state ensembles} $\{p_i,\psi_i\}$, i.e. different convex decompositions of $\rho$ into pure states: $\varrho=\sum_ip_i \psi_i$,  
$\sum_ip_i=1$,  $p_i \geq 0$ (for computation techniques of convex roofs see e.g. \cite{Toth-2015}).
To specify $f$ one can 
take a real, non-negative, LU invariant function, satisfying $f(a|\psi\rangle)=|a|^{2}f(|\psi\rangle)$ as well as $f(a|\psi\rangle \otimes |\phi_{A_{k}}\rangle + 
b |\psi'\rangle \otimes |\phi'_{A_{k}}\rangle) \leq
|a|^{2}f(|\psi\rangle \otimes |\phi_{A_{k}}\rangle) + 
|b|^{2} f(|\psi'\rangle \otimes |\phi'_{A_{k}}\rangle)$, for orthonormal pairs of vectors  $\{ |\phi_{A_{k}}\rangle, |\phi'_{A_{k}}\rangle \}$,  $k=1,\ldots, N$.
Such a choice allows one \cite{DD-et-al-06} 
to select a subset of non-negative parameters $\{ p_{s_1 \ldots s_N} \}$, for which one can 
construct a
convex-roof strong entanglement monotone 
starting from 
 pure states called {\it multipartite concurrence} \cite{Mitert-et-al-05}:
$C_{\cal A}(|\psi\rangle)=2\sqrt{\langle \psi|\otimes \langle \psi| {\cal A} |\psi\rangle \otimes |\psi\rangle}$ with ${\cal A}=\sum_{s_1 \ldots s_N} p_{s_1 \ldots s_N}P^{(s_1)}_{1} \otimes \ldots P^{(s_N)}_{N}$, $s_i=\pm 1$, $P^{(\pm 1)}_{k}$ --- projector onto symmetric (antisymmetric) subspace of $d \otimes d$ "doubled" local space ${\cal H}_{A_{k} \tilde{A}_{k}}$. 
 Any function $f$ invariant under SLOCC filtering, 
 (which technically means that it is constant on G-orbits of pure states,
 i.e. $f(\otimes _i L_i |\psi_{A_1 \cdots A_N}\rangle)=f(|\psi_{A_1 \cdots A_N}\rangle)$, for  $\otimes _i L_i \in G$), satisfies the strong monotonicity criterion (i') for fixed dimensions (see \cite{Verstraete-2003,Gour-2024}).
 The same property concerns the measures based on hyperdeterminants \cite{Mi04}. Sometimes, one may drop the dimension condition, like in the case of concurrence which is a bipartite variant of $C_{\cal A}$ above, but it requires an extra proof --- see \cite{DD-et-al-06}. One of the important multipartite entanglement measures is  {\it the geometric measure of entanglement} \cite{Geom1,Geom2} which is a convex roof extension of pure state measure  (see \cite{Shimony}) $f(\psi)=1 - \mathrm{max}_{\phi} |\langle \phi| \psi \rangle |^{2}$ with maximum taken over all  nonentangled (separable) states. It should be noted that a useful tool for multipartite entanglement is a {\it mixed convex roof} involving mixed states ensembles 
 -- see \cite{PazSilva-Reina-2009,Yang-Et-Al-2009} and references therein. 
\medskip 

For the details of procedures of detection of multipartite entanglement  
the reader is advised to consult Ref. \cite{Guehne-Toth-2009,Nature-Phys-Review}, 
in which 
linear  and nonlinear entanglement witnesses are discussed. 
Several such techniques have already 
been implemented in laboratories.

\section{Final remarks}

The 1933 Nobel Prize in Physics went to 
E.~Schr{\"o}dinger and P.~A.~M.~Dirac for their work on quantum mechanics.
In 2022 this prize was awarded to 
A. Aspect, J. F. Clauser, and A. Zeilinger 
for {\sl experiments with entangled photons, establishing the violation of Bell inequalities and pioneering quantum information science}. According to a broad understanding, the latter prize moves quantum information to a phase in which its applied component becomes as relevant as basic research. Multipartite entanglement is not an exception, 
as several its applications in quantum information processing have been proposed. 

\bigskip

{\bf Acknowledgements}. It is a pleasure to thank Jakub Czartowski for preparing Fig. 2 and several useful remarks. 
We are very grateful to
Otfried G{\"u}hne,
Karol Horodecki, Marcus Huber and Adam Sawicki
for critical reading of the manuscript
and numerous suggestions 
which allowed us to improve of the text.


\end{document}